\documentclass{emulateapj}
\usepackage{graphicx}
\usepackage{subfigure}
\newcommand{\secpoint}{\mbox{$''\mskip-7.6mu.\,$}}

\slugcomment{Received DATE; accepted DATE}

\begin{document}

\title{Rest-Frame Optical Spectra of Three Strongly 
Lensed Galaxies At z$\sim$2\altaffilmark{1}}

\shorttitle{LENSED GALAXIES AT Z$\sim$2}
\shortauthors{HAINLINE ET AL.}

\author{\sc Kevin N. Hainline, Alice E. Shapley\altaffilmark{2,3}, 
and Katherine A. Kornei}
\affil{Department of Astronomy, University of California,
Los Angeles, 430 Portola Plaza, Los Angeles, CA 90024}

\author{\sc Max Pettini}
\affil{Institute of Astronomy, University of Cambridge, 
Madingley Road, Cambridge CB3 0HA, UK}

\and

\author{\sc Elizabeth Buckley-Geer, Sahar S. Allam, and Douglas L. Tucker}
\affil{Fermi National Accelerator Laboratory, Batavia, IL 60510}

\author{}

\altaffiltext{1}{Based, in part, on data obtained at the W.M. Keck
Observatory, which is operated as a scientific partnership among the
California Institute of Technology, the University of California, and
NASA, and was made possible by the generous financial support of the W.M.
Keck Foundation.}
\altaffiltext{2}{Alfred P. Sloan Fellow}
\altaffiltext{3}{David and Lucile Packard Fellow}

\begin{abstract}

We present Keck II NIRSPEC rest-frame optical spectra for three recently 
discovered lensed galaxies: the Cosmic Horseshoe ($z = 2.38$), the Clone ($z = 
2.00$), and SDSS J090122.37+181432.3 ($z = 2.26$). The boost in signal-to-noise 
ratio (S/N) from gravitational lensing provides an unusually detailed view of 
the physical conditions in these objects. A full complement of high S/N 
rest-frame optical emission lines is measured, spanning from rest-frame 3600 to 
6800 \AA, including robust detections of fainter lines such as H$\gamma$, 
[\ion{S}{2}]$\lambda$6717,6732, and in one instance [\ion{Ne}{3}]$\lambda$3869. 
SDSS J090122.37+181432.3 shows evidence for AGN activity, and therefore we focus 
our analysis on star-forming regions in the Cosmic Horseshoe and the Clone. For 
these two objects, we estimate a wide range of physical properties. Current 
lensing models for the Cosmic Horseshoe and the Clone allow us to correct the 
measured H$\alpha$ luminosity and calculated star-formation rate (SFR). 
Metallicities have been estimated with a variety of indicators, which span a 
range of values of $12+$log(O/H) = $8.3 - 8.8$, between $\sim 0.4$ and $\sim 
1.5$ of the solar oxygen abundance. Dynamical masses were computed from the 
H$\alpha$ velocity dispersions and measured half-light radii of the 
reconstructed sources. A comparison of the Balmer lines enabled measurement of 
dust reddening coefficients. Variations in the line ratios between the different 
lensed images are also observed, indicating that the spectra are probing 
different regions of the lensed galaxies. In all respects, the lensed objects 
appear fairly typical of UV-selected star-forming galaxies at $z\sim2$. The 
Clone occupies a position on the emission-line diagnostic diagram of 
[\ion{O}{3}]/H$\beta$ vs. [\ion{N}{2}]/H$\alpha$ that is offset from the 
locations of $z\sim0$ galaxies. Our new NIRSPEC measurements may provide 
quantitative insights into why high-redshift objects display such properties. 
From the [\ion{S}{2}] line ratio, high electron densities ($\sim$1000 cm$^{-3}$) 
are inferred compared to local galaxies, and [\ion{O}{3}]/[\ion{O}{2}] line 
ratios indicate higher ionization parameters compared to the local population. 
Building on previous similar results at $z\sim2$, these measurements provide 
further evidence (at high S/N) that star-forming regions are significantly 
different in high-redshift galaxies, compared to their local counterparts.
\end{abstract}


\section{Introduction}

A wealth of information about star-forming galaxies is contained in their 
optical spectra. The ratios of the fluxes of optical emission lines can be used 
to understand the physical conditions of the gas and stars in star-forming 
regions, including metallicities, temperatures, densities, and ionization 
parameters. Near-infrared spectroscopy is used to study these properties in 
high-redshift galaxies, as the strong rest-frame optical lines are
redshifted into the near-IR at $z > 1.5$. These lines include 
[\ion{O}{2}]$\lambda$3727, H$\beta$, [\ion{O}{3}]$\lambda$$\lambda$5007,4959, 
H$\alpha$, [\ion{N}{2}]$\lambda$6584, and 
[\ion{S}{2}]$\lambda$$\lambda$6717,6731, and can be used to infer the oxygen 
abundance, electron density and ionization parameter in \ion{H}{2} regions. 

Results from a small existing sample indicate that $z\sim2$ galaxies have 
intrinsically different rest-frame optical line ratios from those in the bulk of 
nearby \ion{H}{2} regions and Sloan Digital Sky Survey (SDSS) star-forming 
galaxies \citep{erb2006a,shapley2005,liu2008}. This difference appears in the 
emission-line diagnostic diagram of [\ion{O}{3}]/H$\beta$ vs. 
[\ion{N}{2}]/H$\alpha$, where high-redshift star-forming regions are found 
offset from an extremely tight sequence formed by local objects 
\citep{kauffmann2003b}. The offset may be caused by differences in the SFR, 
average ionization parameter, shape of the stellar ionizing spectrum, \ion{H}{2} 
electron density, or the increased contribution of an AGN or shocked gas to the 
integrated spectra. Based on the analysis of SDSS local star-forming galaxies,
\citet{brinchmann2008} suggest that an elevated ionization parameter
due to high electron densities and non-zero escape fractions of hydrogen
ionizing photons is the primary factor.
Specifying the cause of this offset using direct measurements
from high-redshift galaxies will provide new insight 
into the physical conditions and chemical abundances of the gas in which the 
stars were being formed during the epoch when the star-formation activity of the 
Universe was at its peak \citep{madau1996,steidel1999,bouwens2007}.

The optimal redshift range for this type of analysis using ground-based near-IR 
spectroscopy is $z=2.0 - 2.5$. In this range, the largest number of strong and 
weak nebular emission lines falls within the windows of atmospheric transmission 
and bluewards of the bright thermal IR background at $\lambda \geq 2.35\mu $m. 
However, $z\sim2$ UV-selected galaxies typically have faint magnitudes 
($K_{Vega} > 20.0$ mag) and small angular sizes ($\leq 1^{\prime\prime}$). Thus, 
it is difficult to obtain the high S/N spectra that are necessary for a truly 
detailed view of high-redshift star-forming regions using anything but long 
integration times with the largest telescope apertures. This problem can be 
addressed by taking advantage of the boost provided by gravitational lensing, 
which magnifies and offers more sensitive views of distant galaxies.

The best-studied high-redshift lensed galaxy is MS 1512-cB58 
\citep[$z=2.73$,][]{yee1996}, a typical $\sim L^*$ Lyman break galaxy (LBG). The 
high magnification \citep[$\sim30$,][]{willlew1996,seitz1998} of cB58 has 
allowed for unusually sensitive analyses of the galaxy's rest-frame UV 
\citep{pettini2000, pettini2002}, optical \citep{teplitz2000}, IR 
\citep{siana2008}, and far-IR \citep{sawicki2001, baker2001} spectra and 
photometry. In particular, rest-frame optical spectra were used to calculate 
dust extinction, virial mass, chemical abundances, and SFRs.  Recently, new 
search techniques for strongly-lensed high-redshift galaxies have yielded 
additional candidates which have been spectroscopically confirmed to lie at $z > 
2$ \citep{allam2007, smail2007}. Three of these candidates, SDSS 
J090122.37+181432.3 ($z = 2.26$, hereafter referred to as SDSS J0901+1814; Diehl 
et al. 2009, in preparation), the Cosmic Horseshoe \citep[$z = 
2.38$;][]{belokurov2007}, and the Clone \citep[$z = 2.00$;][]{lin2008} are at 
redshifts even better-suited to rest-frame optical spectroscopy than cB58. 
Analysis of the rest-frame optical spectra of these objects presents a unique 
opportunity to study both \ion{H}{2} region physics and the difference between 
local star-forming galaxies and those at high-redshift.

In this paper, we analyze Keck II NIRSPEC spectra for these three lensed 
objects. In \S \ref{sec:obsdata} we describe the observations, data reduction, 
and line flux measurements. We discuss several physical quantities derived from 
these line fluxes in \S \ref{sec:physquant}. We investigate how our objects 
compare to other objects at $z\sim2$ in \S \ref{sec:analysis}, and review the 
physical quantities for these objects in order to explain the differences in the 
physical conditions between high-redshift and local \ion{H}{2} regions. Finally, 
we conclude in \S \ref{sec:conclusions}. Throughout, we assume a cosmology with 
$\Omega_{M} = 0.27$, $\Omega_{\Lambda} = 0.73$, and $H_0 = 71$ km s$^{-1}$ 
Mpc$^{-1}$.

\section{Observations and Data Reduction}\label{sec:obsdata}

\subsection{NIRSPEC Observations\label{sec:nirspecobs}}

\begin{figure*}
\epsscale{1.} 
\plotone{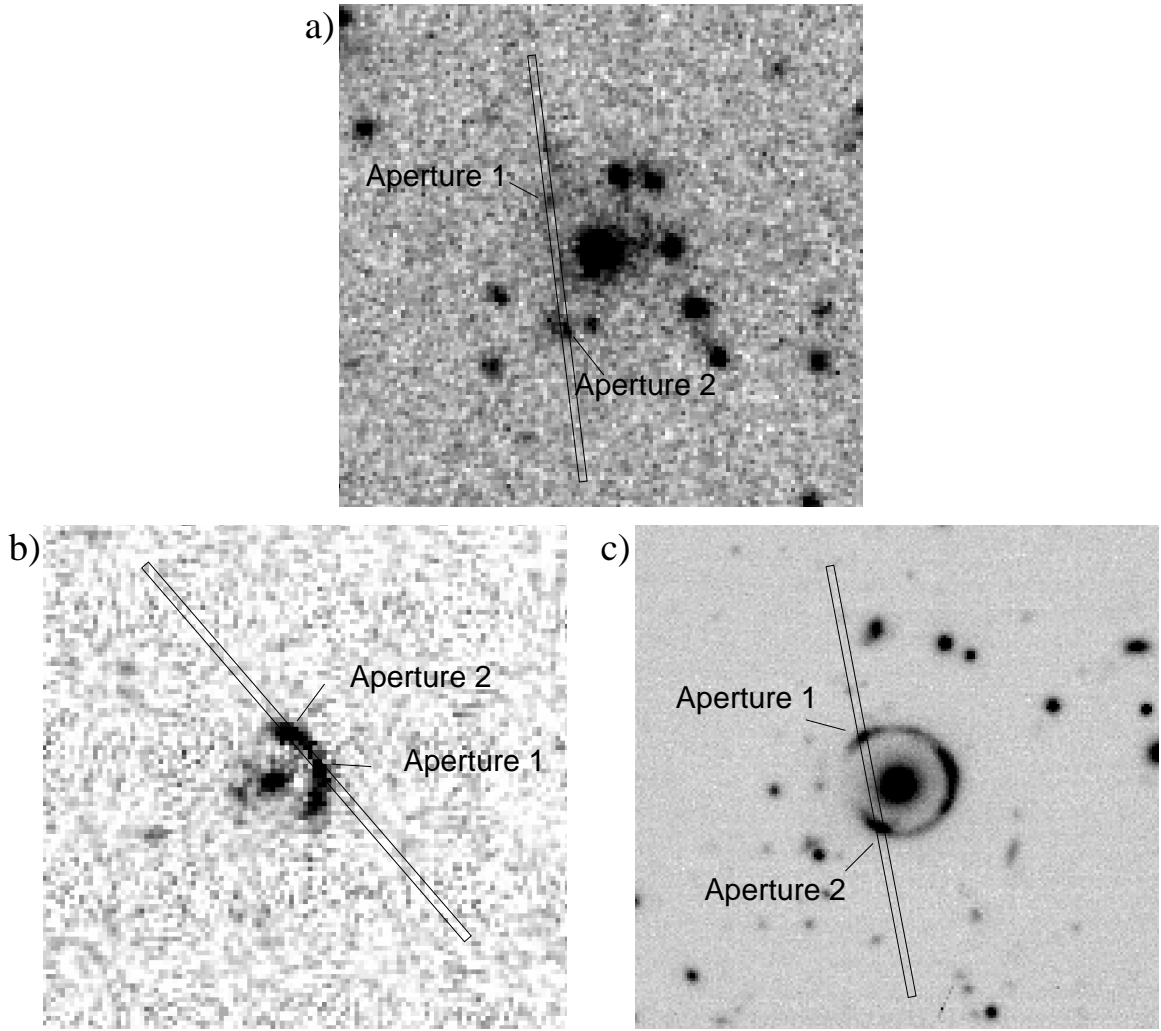} 
\caption{Images of the Objects with the NIRSPEC slit overlaid, defining the 
various apertures used in the paper. a) SDSS J0901+1814, from a SDSS $r$-band 
image, b) The Clone, from a SDSS $g$-band image, and c) The Cosmic Horseshoe, 
from a VLT/FORS2 $R$-band image (Lindsay King, private communication). The 
NIRSPEC slit shown is 0\secpoint76 $\times 42^{\prime\prime}$, and the 
separation between the different apertures are 12\secpoint52 for SDSS 
J0901+1814, 3\secpoint95 for the Clone.\label{fig:spec1}, and 8\secpoint82 for 
the Cosmic Horseshoe.}
\epsscale{1.}
\end{figure*}

We obtained near-IR spectra on 2008 February 19 using the NIRSPEC spectrograph 
\citep{mclean98} on the Keck~II telescope. Our targets included three strongly 
lensed high redshift galaxies: SDSS J090122.37, the Cosmic Horseshoe, and the 
Clone \citep[Diehl et al. 2009;][]{belokurov2007,lin2008}. At $z\sim2$, several 
strong rest-frame optical lines fall within windows of atmospheric transmission 
in the near-IR. H$\alpha$ and [\ion{N}{2}] fall within the NIRSPEC 6 (similar to 
$K$-band) filter for all three objects. H$\gamma$, [\ion{O}{3}]$\lambda$4363, 
H$\beta$, and [\ion{O}{3}]$\lambda$$\lambda$4959,5007 fall in the NIRSPEC 5 
(similar to $H$-band) filter for the Cosmic Horseshoe and SDSS J090122.37, and 
NIRSPEC 4 for the Clone. [\ion{O}{2}] and [\ion{Ne}{3}] fall in the NIRSPEC 3 
(similar to $J$-band) filter for the Cosmic Horseshoe and SDSS J090122.37, and 
NIRSPEC 2 for the Clone. Exposure times for the different filters ranged from 
$3\times600$ s to $9\times600$ s, and are listed in Table \ref{tab:obs}. All 
targets were observed with a 0\secpoint76 $\times$ $42^{\prime\prime}$ long 
slit. Conditions were photometric during the night and seeing ranged from 
0\secpoint4 - 0\secpoint9. The spectral resolution as determined from sky lines 
was $\sim10$\,\AA\, for the NIRSPEC-2, -3, -4, and -5 filters, and 
$\sim15$\,\AA\, for the NIRSPEC-6 filter. For each object, we placed the slit 
across two of the lensed images to look for variation in the line ratios between 
them. The slit position angles (in degrees east of north) were $7.22^{\circ}$ 
for SDSS J090122.37, $10.73^{\circ}$ for the Cosmic Horseshoe, and 
$219.97^{\circ}$ for the Clone, and were determined by the locations of high 
surface brightness knots in the optical lensed images. The separation between 
the different apertures are 12\secpoint52 for SDSS J090122.37, 8\secpoint82 for 
the Cosmic Horseshoe, and 3\secpoint95 for the Clone. Images of the objects with 
slits overlaid are presented in Figure \ref{fig:spec1}.

\begin{deluxetable*}{lccccccc}
\tablewidth{0pt} \tabletypesize{\footnotesize}
\tablecaption{Galaxies Observed with Keck~II
NIRSPEC\label{tab:obs}}
\tablehead{
\colhead{~~~~~~NAME~~~~~~} &
\colhead{R.A. (J2000)} &
\colhead{Dec. (J2000)} &
\colhead{$z_{{\rm H}\alpha}$} &
\colhead{~$r (mag)$~}  &
\colhead{$g-r$$(mag)^a$} &
\colhead{Exposure (s)} &
\colhead{Band}
}
\startdata
SDSS J0901+1814 \dotfill  &  09 01 22.37  &  18 14 32.35 &  2.2586  & 20.6 & 0.52, 0.21 & 3 $\times$ 600 & $J$ \\
 & & & & & & 6 $\times$ 600 & $H$ \\
 & & & & & & 6 $\times$ 600 & $K$ \\
Cosmic Horseshoe \dotfill   &  11 48 33.14  &  19 30 3.20  &  2.3813 &  19.0 & 0.29, 0.25 & 3 $\times$ 600 & $J$ \\
 & & & & & & 9 $\times$ 600 & $H$ \\
 & & & & & & 8 $\times$ 600 & $K$ \\
 Clone \dotfill  &  12 06 2.09  &  51 42 29.52 &  2.0026  &  19.0 & 0.39, 0.30, & 3 $\times$ 600 & $J$ \\
 & & & & & 0.34 & 3 $\times$ 600 & $H$ \\
 & & & & & & 3 $\times$ 600 & $K$ \\
 
\enddata
\tablecomments{Units of right ascension are hours, minutes, and seconds, and units of declination are degrees, arcminutes and arcseconds.}
\tablenotetext{a}{These colors are listed for apertures 1, 2, and in the case of the Clone, the full aperture as shown in Figure \ref{fig:spec1}.}
\end{deluxetable*}

\subsection{Data Reduction and Optimal Background Subtraction\label{sec:dataredt}}

Data reduction was performed following the procedure described in 
\citet{liu2008}, where the sky background was subtracted using an optimal method 
on the two-dimensional spectral images \citep[][Becker, private 
communication]{kelson03}. One notable exception to this procedure addressed the 
matter of pattern noise. A large fraction of the exposures taken in 2008 
February were affected by pattern noise, which consisted of a constant positive 
or negative offset in the mean count level in every $8^{\mathrm{th}}$ row of the 
upper-right hand quadrant of the NIRSPEC CCD images. Prior to the standard 
reduction procedure, we removed this pattern noise in all 75 of the science 
images.

Sky subtraction for the Cosmic Horseshoe observations required extra care, due 
to the presence of continuum emission in the slit from the outskirts of the 
lensing galaxy \citep[a luminous red galaxy at $z=0.444$;][]{belokurov2007}. 
Indeed, the first stage of sky subtraction typically consisted of a simple 
difference of adjacent, dithered exposures. The dithers adopted for this target 
were (unfortunately) similar in size to the spacing between the positions of the 
Horseshoe and lens images, leading to the subtraction of the lens continuum 
(rather than blank sky) from the Horseshoe in some pairs of exposures. We 
quantified and corrected for this effect using pairs of exposures in which no 
oversubtraction occurred.

\begin{figure*}
\epsscale{1.} 
\plotone{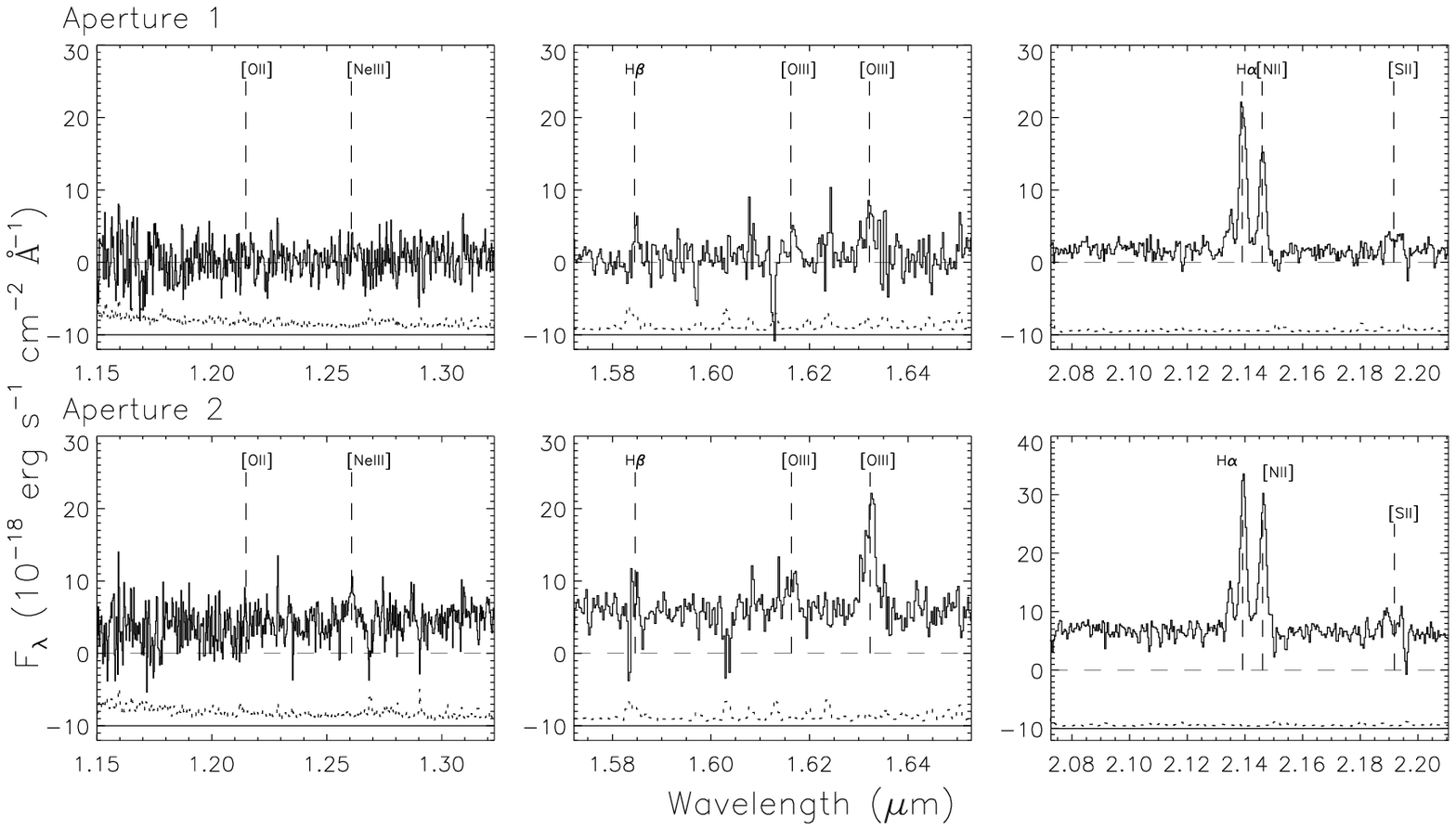} 
\caption{NIRSPEC spectrum of SDSS J0901+1814. The top row contains the $J$, $H$, 
and $K$-band spectra for aperture 1, while the bottom row shows those for 
aperture 2. Apertures 1 and 2 are as labelled in Figure \ref{fig:spec1}. The 
positions of the prominent emission lines are indicated. The dotted lower 
spectrum indicates the 1$\sigma$ errors offset vertically by $-10\times10^{-18}$ 
erg s$^{-1}$ cm$^{-2}$ \AA$^{-1}$ for clarity. Larger errors occur at the 
position of the sky lines and at points of high atmospheric extinction.
\label{fig:spec3817}} 
\epsscale{1.}
\end{figure*}

\begin{figure*}
\epsscale{1.} 
\plotone{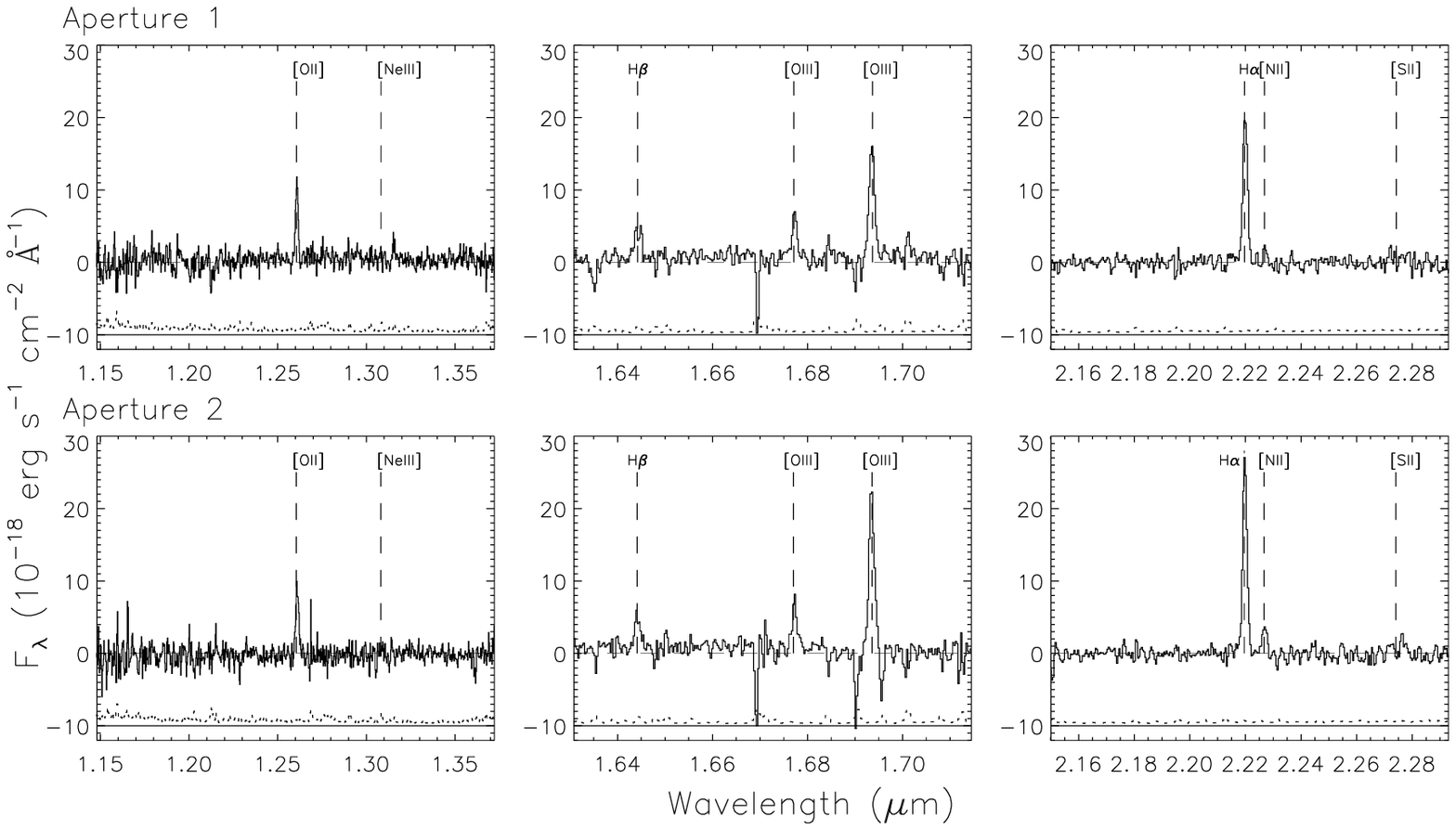} 
\caption{NIRSPEC spectrum of the Cosmic Horseshoe. The top row contains the $J$, 
$H$, and $K$-band spectra for aperture 1, while the bottom row shows those for 
aperture 2. Apertures 1 and 2 are as labelled in Figure \ref{fig:spec1}. The 
positions of the prominent emission lines are indicated. The dotted lower 
spectrum indicates the 1$\sigma$ errors offset vertically by $-10\times10^{-18}$ 
erg s$^{-1}$ cm$^{-2}$ \AA$^{-1}$ for clarity. Larger errors occur at the 
position of the sky lines and at points of high atmospheric extinction.
\label{fig:specCH}} 
\epsscale{1.}
\end{figure*}

\begin{figure*}
\epsscale{1.} 
\plotone{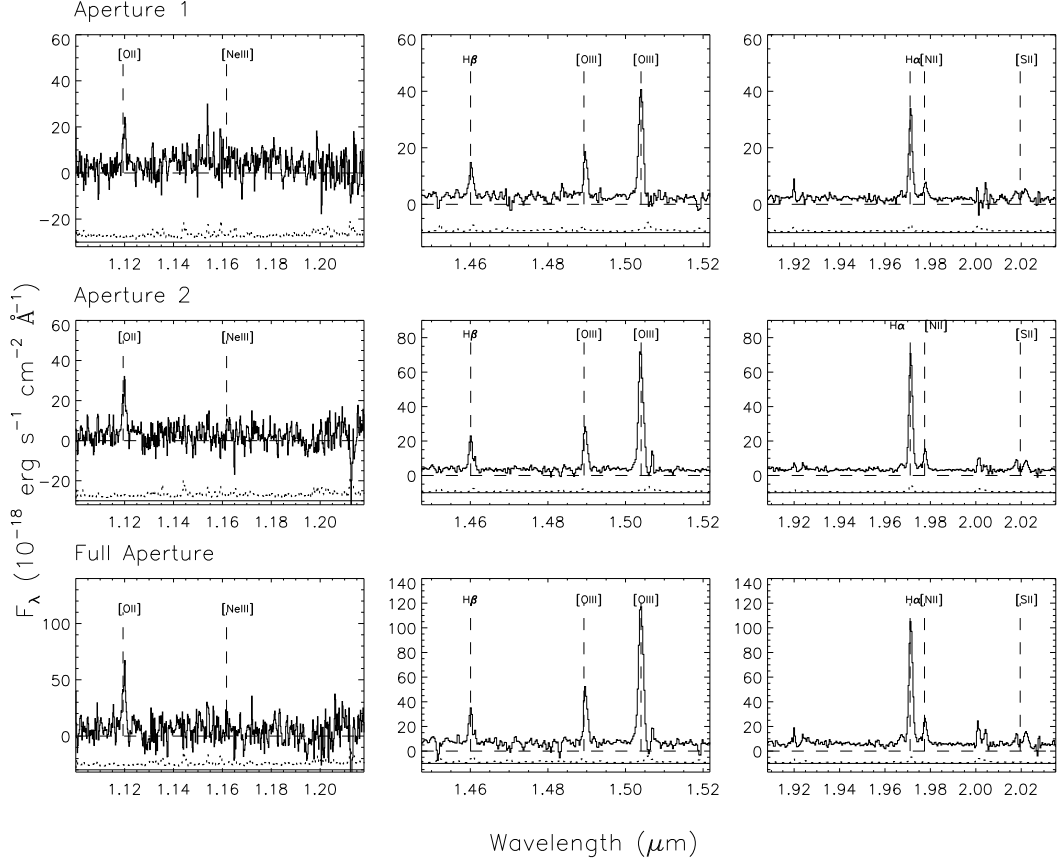} 
\caption{NIRSPEC spectrum of the Clone. The top row contains the $J$, $H$, and 
$K$-band spectra for aperture 1, the second row shows those for aperture 2, and 
the bottom row shows those from the full aperture. Apertures 1 and 2 are as 
labelled in Figure \ref{fig:spec1}. The positions of the prominent emission 
lines are indicated. The dotted lower spectrum indicates the 1$\sigma$ errors 
offset vertically by $-30\times10^{-18}$ erg s$^{-1}$ cm$^{-2}$ \AA$^{-1}$ for 
the $J$-band spectra, and $-10\times10^{-18}$ erg s$^{-1}$ cm$^{-2}$ \AA$^{-1}$ 
for the $H$- and $K$-band spectra for clarity. Larger errors occur at the 
position of the sky lines and at points of high atmospheric extinction.
\label{fig:specClone}}
\epsscale{1.}
\end{figure*}

\begin{figure*}
\epsscale{1.} 
\plotone{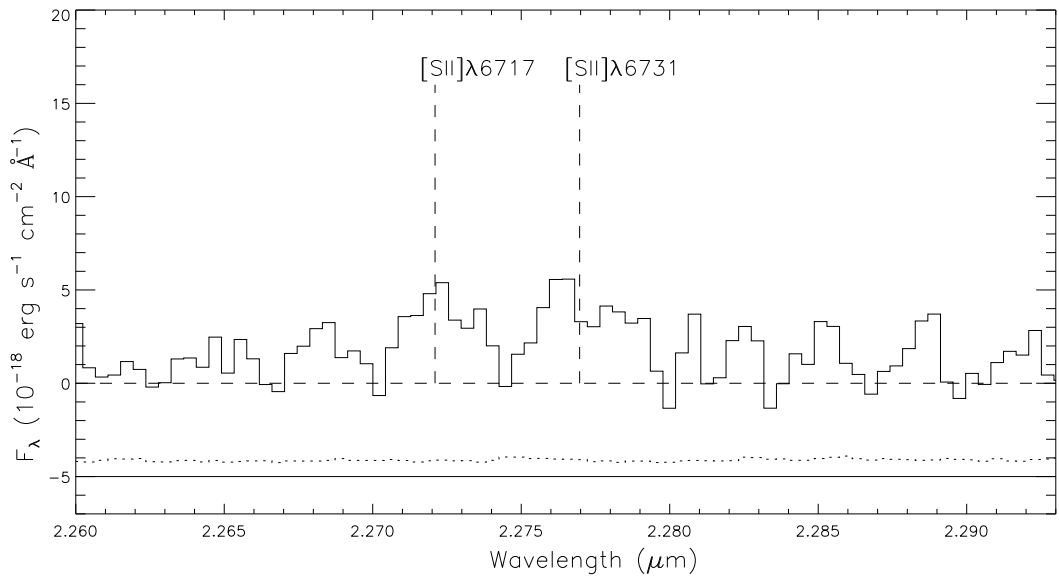} 
\caption{NIRSPEC summed spectra from both Cosmic Horseshoe apertures, for the 
[SII] doublet. The dotted lower spectrum indicates the 1$\sigma$ errors offset 
vertically by $-5\times10^{-18}$ erg s$^{-1}$ cm$^{-2}$ \AA$^{-1}$ for clarity. 
\label{fig:specCHSII}}
\epsscale{1.}
\end{figure*}

One-dimensional spectra were extracted from these two-dimensional reduced images 
along with error spectra. For all three of our objects we extracted apertures 
for each of the two lensed images covered. For the Clone, the lensed objects 
almost overlapped, so we extracted a third, full aperture that extended along 
the slit across both lensed images. These spectra were then flux-calibrated 
using A-type star observations according to the method described in 
\citet{shapley2005} and \citet{erb2003}.

\subsection{Line Flux Measurements\label{sec:lineflux}}

\begin{figure*}
\epsscale{1.} 
\plotone{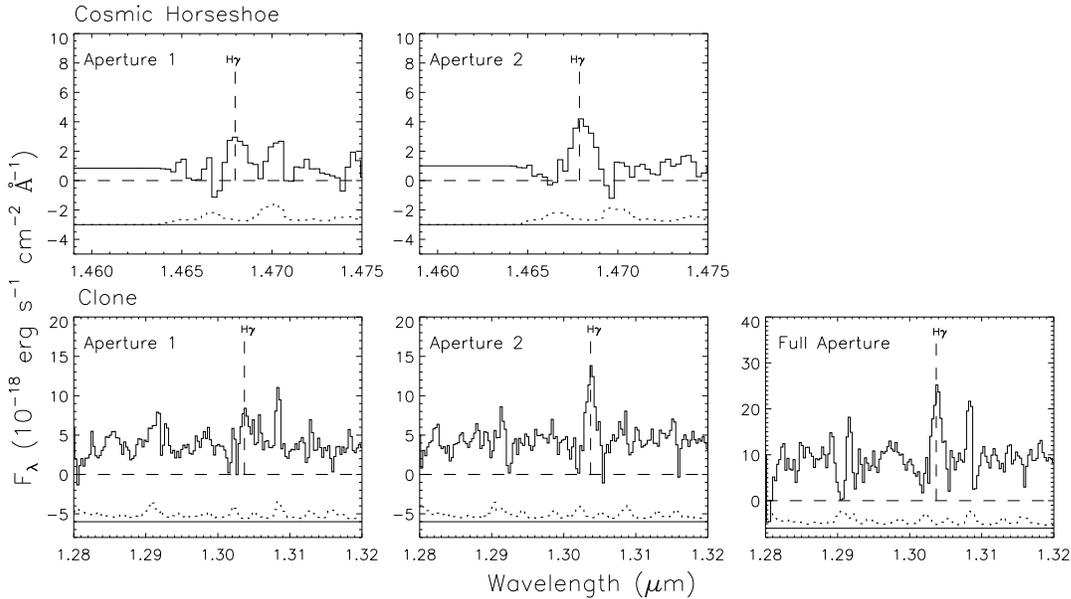} 	 
\caption{NIRSPEC spectra of the H$\gamma$ lines for the target objects. The 
dotted lower spectra indicate the 1$\sigma$ errors offset vertically by 
$-5\times10^{-18}$ erg s$^{-1}$ cm$^{-2}$ \AA$^{-1}$ for 
clarity.\label{fig:Hgamma}}
\epsscale{1.}
\end{figure*}

One-dimensional, flux-calibrated spectra are shown along with the error spectra 
in Figures \ref{fig:spec3817}, \ref{fig:specCH}, and \ref{fig:specClone}. The 
error spectrum represents the 1$\sigma$ uncertainty in the flux at each 
wavelength. In addition, zoomed-in regions of select portions of the spectra are 
featured in Figures \ref{fig:specCHSII} and \ref{fig:Hgamma}. The redshifts of 
the lensed objects were measured by fitting a Gaussian profile to the H$\alpha$ 
emission feature. In all cases, the redshifts from the separate components of 
the lensed image pairs agreed to within $\Delta z/(1+z)$ = 0.0001 ($\Delta$$v$ = 
$27 - 30$ km s$^{-1}$).

The lines that we set out to measure were 
[\ion{O}{2}]$\lambda$$\lambda$3726,3729, [\ion{Ne}{3}]$\lambda$3869, H$\gamma$, 
[\ion{O}{3}]$\lambda$4363, H$\beta$, [\ion{O}{3}]$\lambda$$\lambda$4959,5007, 
H$\alpha$, [\ion{N}{2}]$\lambda$6584, and 
[\ion{S}{2}]$\lambda$$\lambda$6717,6732. Line fluxes were determined by fitting 
Gaussian profiles using the IRAF task, SPLOT. The H$\beta$, H$\gamma$, and 
[\ion{O}{2}]$\lambda$$\lambda$3726,3729 lines were measured individually, while 
the pairs of [\ion{O}{3}]$\lambda$$\lambda$4959,5007, H$\alpha$ and 
[\ion{N}{2}]$\lambda$6584, and [\ion{S}{2}]$\lambda$$\lambda$6717,6732 were 
fitted simultaneously. With this method, we fixed the central wavelength and 
FWHM based on the best-fit parameters for the brighter line and then obtained a 
combined fit for the relative fluxes of the two lines.

In some cases, we had to measure specific lines through different methods 
because of the presence of sky lines, or the weak fluxes of the lines. For the 
Cosmic Horseshoe and, to a lesser extent, the Clone, sky systematics were an 
issue. In both objects, the H$\beta$ line was detected, but its wavelength 
coincided with that of a strong sky line. Accordingly, we treated the measured 
H$\beta$ line flux as a lower limit. In \S \ref{sec:physquant}, we describe a 
method for recovering the actual H$\beta$ flux in the face of sky systematics. 
We measured a lower limit on the line flux for the H$\beta$ and 
[\ion{S}{2}]$\lambda$6732 features of aperture 2 of SDSS J0901+1814 as these 
lines were over-subtracted on one side during the sky subtraction process. We 
placed upper limits on the [\ion{O}{3}]$\lambda$4363 line fluxes in all 
apertures, and on the [\ion{Ne}{3}]$\lambda$3869 line fluxes in all but aperture 
1 of SDSS J0901+1814 where it was detected. We summed the flux across both of 
the [\ion{S}{2}] lines for aperture 1 of both SDSS J0901+1814 and the Cosmic 
Horseshoe due to the low flux from the individual lines.

Uncertainties in line fluxes were estimated using a Monte Carlo approach. For 
each aperture and filter, we generated five hundred artificial spectra. Fake 
spectra were created by perturbing the flux at each wavelength of the true 
spectrum by a random amount consistent with the 1$\sigma$ error spectrum. Line 
fluxes were measured from these simulated spectra using the same procedure that 
was applied to the actual data. The standard deviation of the distribution of 
line fluxes measured from the artificial spectra was adopted as the error on 
each line flux measurement. Emission-line fluxes and associated uncertainties 
are given in Table \ref{tab:emi}.

\begin{deluxetable*}{cccccccc}
\tabletypesize{\scriptsize}
\tablecaption{Emission Line Fluxes\label{tab:emi}$^a$}
\tablewidth{0pt}
\tablehead{
\colhead{} & \colhead{SDSS J0901+1814} & \colhead{SDSS J0901+1814} & \colhead{Cosmic Horseshoe} & \colhead{Cosmic Horseshoe} & \colhead{Clone} & \colhead{Clone} & \colhead{Clone}
}
\startdata

Aperture & 1 & 2 & 1 & 2 & 1 & 2 & Full \\
${\sigma_{\mathrm{H}\alpha}}^b$ & 131$\pm$5 &  112$\pm$4 & 69$\pm$4 & 58$\pm$3 & 71$\pm$8 & 73$\pm$6 & 80$\pm$4 \\
$\mathrm{F}_{[\mathrm{OII}]}$ & $<$9.0 & $<$10.0 & 18.8$\pm$0.9 & 20.0$\pm$1.0 & 29.0$\pm$3.0 & 47.0$\pm$3.0 & 85.0$\pm$5.0 \\
$\mathrm{F}_{[\mathrm{NeIII}]\lambda3869}$ & $<$8.0 & 19.0$\pm$3.0 & $<$4.0 & $<$4.0 & $<$17.0 & $<$16.0 & $<$28.0 \\
$\mathrm{F}_{[\mathrm{H}\gamma]}$ & - & - & 3.0$\pm$1.0 &  4.6$\pm$0.4 & 6.0$\pm$2.0 & 11.2$\pm$0.7 & 21.0$\pm$1.0 \\
$\mathrm{F}_{[\mathrm{OIII}]\lambda4363}$ & - & - & $<$1.5  & $<$1.6 & $<$2.6 & $<$2.4 & $<$4.3 \\
$\mathrm{F}_{\mathrm{H}\beta}$ & 5.0$\pm$2.0 & $>$$4^{c}$ & 8.0$\pm$1.0 & 7.0$\pm$1.0 & 15.0$\pm$2.0 & 25.0$\pm$2.0 & 33.0$\pm$4.0 \\
$\mathrm{F}_{\mathrm{H}\beta,\mathrm{inf}^d}$ & - & - & 12.6$\pm$0.3 & 16.2$\pm$0.3 & 15.4$\pm$0.6 & 36.0$\pm$1.0 & 53.0$\pm$1.2 \\
$\mathrm{F}_{[\mathrm{OIII}]\lambda4959}$ & 8.0$\pm$2.0 & 11.0$\pm$2.0 & 7.5$\pm$0.8 & 9.6$\pm$0.5 & 20.0$\pm$1.0 & 35.0$\pm$1.0 & 58.0$\pm$2.0 \\
$\mathrm{F}_{[\mathrm{OIII}]\lambda5007}$ & 18.0$\pm$4.0 & 39.0$\pm$3.0 & 21.9$\pm$0.6 & 31.2$\pm$0.6 & 56.0$\pm$1.0 & 106.0$\pm$1.0 & 173.0$\pm$2.0 \\
$\mathrm{F}_{\mathrm{H}\alpha}$ & 63.0$\pm$2.0 & 72.0$\pm$1.0 & 43.4$\pm$0.9 & 53.9$\pm$0.9 & 61.0$\pm$2.0 & 132.0$\pm$4.0 & 202.0$\pm$5.0 \\
$\mathrm{F}_{[\mathrm{NII}]\lambda6584}$ & 42.0$\pm$1.0 & 64.0$\pm$1.0 & 4.0$\pm$0.7 & 8.7$\pm$0.7 & 10.8$\pm$0.9 & 21.0$\pm$1.0 & 39.0$\pm$2.0 \\
$\mathrm{F}_{[\mathrm{SII}]\lambda6717}$ & 12.0$\pm$1.0$^{e}$ & 11.0$\pm$1.0 & 6.0$\pm$$2.0^{e,g}$ & 4.0$\pm$$1.0^{g}$ & 5.5$\pm$0.9 & 10.0$\pm$1.0 & 9.0$\pm$2.0 \\
$\mathrm{F}_{[\mathrm{SII}]\lambda6732}$ & 12.0$\pm$1.0$^{e}$ & $>$$6.4^{f}$ & 6.0$\pm$$2.0^{e,g}$ & 6.0$\pm$1.0$^{g}$ & 7.3$\pm$0.9 & 10.0$\pm$1.0 & 12.0$\pm$3.0 \\

\enddata

\tablenotetext{a}{Emission line flux and error in units of $10^{-17}$ ergs $\mathrm{s}^{-1}$ $\mathrm{cm}^{-2}$}
\tablenotetext{b}{One-dimensional H$\alpha$ velocity dispersion and error in units of km s$^{-1}$. }
\tablenotetext{c}{A sky line coincided with H$\beta$ for aperture 2 of SDSS J0901+1814, and the line flux value was found by summing only the positive flux, instead of fitting the line with a Gaussian.}
\tablenotetext{d}{These inferred H$\beta$ values are calculated from the observed broadband $g-r$ color and the value for H$\alpha$, as described in \S \ref{sec:dust}.}
\tablenotetext{e}{This value corresponds to the sum of the flux across both lines, with error.}
\tablenotetext{f}{A sky line coincided with [SII]$\lambda6732$ for SDSS J0901+1814, aperture 2, and the line flux value was found by summing only the positive flux, instead of fitting the line with a Gaussian.}
\tablenotetext{g}{The spectra from apertures 1 and 2 were summed for the Cosmic Horseshoe in order to better measure the [SII] line fluxes. The measured values were $\mathrm{F}_{[\mathrm{SII}]\lambda6717}$ = $8\pm2$ , and $\mathrm{F}_{[\mathrm{SII}]\lambda6732}$ = $8\pm2$.}
\end{deluxetable*}

\section{Physical Quantities}\label{sec:physquant}

The flux measurement of rest-frame optical emission lines allows us to probe the 
physical state of the interstellar gas in the target objects. Because of the 
magnification provided by gravitational lensing, we have detected lines with a 
higher signal-to-noise ratio (S/N) than what is typically seen in unlensed 
high-redshift objects. We can measure dust extinction from the ratios of the 
Balmer lines of H$\alpha$, H$\beta$, and H$\gamma$, while the H$\alpha$ line 
offers the highest S/N measure of the gas velocity dispersion, as well as the 
SFR of the galaxy. The set of [\ion{O}{3}]$\lambda$$\lambda$4959,5007, 
[\ion{O}{2}]$\lambda$$\lambda$3726,3729, [\ion{N}{2}]$\lambda$6584 and Balmer 
emission lines provides multiple methods of determining the gas-phase oxygen 
abundance. The ratio of [\ion{O}{2}] and [\ion{O}{3}] lines offers a probe of 
the ionization parameter of the star-forming regions. Finally, we can use our 
detection of the [\ion{S}{2}]$\lambda$$\lambda$6717,6732 lines and an upper 
limit of the [\ion{O}{3}]$\lambda$4363 line to estimate the electron density and 
electron temperature of the gas. Table~\ref{tab:phy} lists several
relevant line ratios and the physical quantities derived from them,
as described in the sections below.

\subsection{AGN Contamination}

We measured the ratio of [\ion{N}{2}]/H$\alpha = 0.65 \pm 0.03$ and $0.88 \pm 
0.03$ for the two apertures of SDSS J0901+1814. Such high ratios, especially in 
the second aperture, cannot be produced in star-forming regions from 
photoionization by massive stars \citep{kewley2001a,kauffmann2003b}. We 
therefore infer the presence of an AGN. With insights from the lens model for 
this system, the AGN can be probed on smaller scales than is typically possible 
at $z\sim2$. Such analysis, however, is outside the scope of the current work. 
For our study of the majority of physical quantities, we focus on the Cosmic 
Horseshoe and the Clone, where the line fluxes do not appear to be contaminated 
by the presence of an AGN.

\subsection{Velocity Dispersion}

The velocity dispersion is a measure of the dynamics of the gas bound to the 
galaxy by gravity. We used H$\alpha$ line widths to estimate velocity dispersion 
after correcting for the instrumental resolution. The velocity dispersions were 
calculated by $\sigma$ = FWHM$/2.355 \times \frac{c}{\lambda}$, where the FWHM 
is the full width at half maximum in wavelength after subtraction of the 
instrumental resolution ($14 - 15$ \AA$\,$in the $K$-band) in quadrature. The 
velocity dispersions we measured were $131\pm5$ km s$^{-1}$ and $112\pm4$ km 
s$^{-1}$ for the apertures of SDSS J0901+1814, $69\pm4$ km s$^{-1}$ and $58\pm3$ 
km s$^{-1}$ for the apertures of the Cosmic Horseshoe, and $71\pm8$ km s$^{-1}$, 
$73\pm6$ km s$^{-1}$, and $80\pm4$ km s$^{-1}$ for the two apertures and the 
full aperture of the Clone. Uncertainties in velocity dispersion were estimated 
using the same Monte Carlo approach that was applied to line fluxes.

\subsection{Dust Extinction\label{sec:dust}}

It is important to understand the effects of dust extinction on the various line
fluxes, as reddening by dust is highly wavelength dependent and thus can alter
the observed flux ratios of widely spaced emission lines from their intrinsic
values. Extinction in $z\geq 2$ star-forming galaxies
is commonly estimated from rest-frame UV colors and an
application of the \citet{calzetti2000} starburst
attenuation law. The Calzetti
law appears to provide a fairly accurate description
on average of the reddening and attenuation of the
UV stellar continuum in both nearby
and distant starburst galaxies \citep{reddy2004,reddy2006a}.
The degree of dust extinction in star-forming regions
can also be estimated from the Balmer lines of hydrogren,
because these strong optical lines have
intrinsic ratios that are well described by atomic theory. Under the assumption
of Case B recombination \citep{agn2} and for T = 10,000 K, the Balmer ratios are
set. Any deviations in the observed line ratios are then attributed to dust
extinction. \citet{calzetti2001} demonstrates that, in local
star-forming galaxies, the stellar continuum suffers
less reddening than the ionized gas, expressed as $E(B-V)_{\rm{star}}=0.44 E(B-V)_{\rm{gas}}$.
Furthermore, \citet{calzetti2001} suggests that
the reddening of the Balmer lines in nearby UV-selected starbursts
is better described by a foreground dust distribution and traditional Milky Way extinction
curve \citep{cardelli1989}. On the other hand, \citet{erb2006c} present evidence
that, in $z\sim 2$ star-forming galaxies,
$E(B-V)_{\rm{star}}\simeq E(B-V)_{\rm{gas}}$ and a \citeauthor{calzetti2000} starburst
extinction law applied to both UV-continuum and H$\alpha$ emission lines
gives rise to the best agreement between UV- and H$\alpha$-derived
SFRs. Our new observations of the Cosmic Horseshoe
and the Clone allow for a detailed comparison of different estimates
of extinction in high-redshift star-forming galaxies.

First,  we used
the observed broadband $g-r$ color for our objects and an assumed intrinsic SED
to obtain an estimate of $E(B-V)$. The SED model assumed was a \citet{bc2003}
solar metallicity, constant SFR model with stellar age of 570 Myr, as
\citet{erb2006b} find a median age of 570 Myr for star forming regions of
galaxies at $z\sim2$. We used SDSS $g$ and $r$ magnitudes for the Cosmic
Horseshoe, and in the case of the Clone, the Apache Point Observatory SPIcam $g$
and $r$ magnitudes of the individual apertures presented in \citet{lin2008}. The
$E(B-V)$ values we calculated were 0.16, 0.13 for the Cosmic Horseshoe, and
0.28, 0.21, and 0.24 for the Clone. 

Next, extinction was estimated
based on rest-frame optical emission lines.
In both the Cosmic Horseshoe and the Clone,
we have detected three of the most prominent Balmer lines: H$\alpha$, H$\beta$,
and (because of the magnification afforded by gravitational lensing) H$\gamma$.
Unfortunately, in both objects, H$\beta$ lines lay over prominent sky
lines with systematically negative residuals, yielding in some cases unphysically
low values of H$\beta$/H$\gamma$. \citet{agn2} gives the value of H$\beta$/H$\gamma$ under Case B 
recombination as
2.14. Because reddening will only serve to increase this ratio, our measured
values of less than 2.14 in aperture 2 of the Horseshoe and the full
aperture of the Clone suggest that, in general,  our measured
H$\beta$ values were systematically low as a result of the residual error in
subtraction of the coincident sky line. H$\gamma$ does not appear to
suffer from the same systematic sky residuals as H$\beta$,
and therefore we used a comparison of the observed and
intrinsic ratios of H$\alpha$/H$\gamma$ to infer
$E(B-V)_{\rm{gas}}$, assuming an intrinsic H$\alpha$/H$\gamma$
ratio of 6.12 \citep{agn2}. To increase the S/N of the
H$\gamma$ feature, we combine spectra from individual
apertures of each object to obtain an average
H$\alpha$/H$\gamma$ and $E(B-V)_{\rm{gas}}$. The observed
H$\alpha$/H$\gamma$ ratios imply values of $E(B-V)_{\rm{gas}}$
of $0.45\pm 0.04$ and $0.28\pm 0.04$ for the Cosmic Horseshoe and the
Clone, respectively.

The value of $E(B-V)_{\rm{gas}}$
for the Cosmic Horseshoe is almost three times
higher than that inferred from the rest-frame
UV continuum. While this difference at first appears
to support the prescription from \citet{calzetti2000},
regarding the relative extinction of starlight and ionized gas,
we argue in section~\ref{sec:sfr} that the $g-r$-based
estimate of $E(B-V)$ provides a more robust estimate
of the ionized gas extinction. In contrast,
both $g-r$ and
H$\alpha$/H$\gamma$ provide consistent estimates
of dust extinction in the Clone, i.e., $E(B-V)_{\rm{star}}\simeq E(B-V)_{\rm{gas}}$.
It is also worth noting that, over the wavelength range probed by
our NIRSPEC spectra, the \citet{calzetti2000} and
\citet{cardelli1989} extinction curves yield very similar values
for $E(B-V)_{\rm{gas}}$, as well as extinction-corrected
line ratios. In all subsequent discussion,
we adopt the \citet{calzetti2000} extinction
law and $E(B-V)$ values based on rest-frame
UV colors.

One important application of
our derived $E(B-V)$ values is for
inferring H$\beta$ fluxes, independent
of sky residuals. We follow a similar methodology
to that of \citet{lemoine2003},
and calculate the H$\beta$ flux based on the observed H$\alpha$ flux,
the intrinsic ratio of 2.86 between
H$\alpha$ and H$\beta$, and our estimated
$E(B-V)$ based on $g-r$ colors. Accordingly,
$F_{H\beta}= F_{H\alpha}/2.86\times 10^{-0.4 E(B-V)(\kappa_{\beta}-\kappa_{\alpha})}$. 
In this expression, $\kappa_{\beta}$ and $\kappa_{\alpha}$ refer
to the extinction coefficients in the \citeauthor{calzetti2000} curve at the wavelengths of
H$\beta$ and H$\alpha$, respectively. Inferred H$\beta$ fluxes are listed
in Table~\ref{tab:emi} and used in the calculation of 
physical quantities in subsequent sections.

\begin{deluxetable*}{cccccc}
\tabletypesize{\scriptsize}
\tablecaption{Calculated Physical Values\label{tab:phy}}
\tablewidth{0pt}
\tablehead{
\colhead{} & \colhead{Cosmic Horseshoe} & \colhead{Cosmic Horseshoe} &  
\colhead{Clone} & \colhead{Clone} & \colhead{Clone}
}
\startdata
Aperture & 1 & 2 & 1 & 2 & Full \\
$R_{23}$ & 3.82$\pm$0.13 & 3.78$\pm$0.11 & 6.88$\pm$0.35 &  5.25$\pm$0.18 & 5.97$\pm$0.18 \\
$R_{23}$$^a$ & 4.08$\pm$0.15 & 3.95$\pm$0.12 & 7.45$\pm$0.40  & 5.52$\pm$0.20 & 6.36$\pm$0.20 \\
$N2$ & -1.04$\pm$0.07 & -0.79$\pm$0.04 & -0.75$\pm$0.04 & -0.79$\pm$0.02  & -0.72$\pm$0.02 \\
$O3N2$ & 1.28$\pm$0.08 & 1.08$\pm$0.04 & 1.31$\pm$0.05 & 1.26$\pm$0.03 &  1.23$\pm$0.02 \\
12+log(O/H)$_{{R}_{23}}$ & 8.85$\pm$0.05 & 8.85$\pm$0.05 & 8.55$ \pm$0.05 & 8.70$\pm$0.05 & 8.63$\pm$0.05 \\
${12+\mathrm{log(O/H)}_{{R}_{23}}}^{a}$ & 8.82$\pm$0.05 & 8.84$ \pm$0.05 & 8.50$\pm$0.05 & 8.68$\pm$0.05 & 8.60$\pm$0.05 \\
12+log(O/H)$_{{{N}2}}$ & 8.31$\pm$0.18 & 8.45$\pm$0.18 & 8.47$ \pm$0.18 & 8.45$\pm$0.18 & 8.50$\pm$0.18 \\
12+log(O/H)$_{{O3N2}}$ & 8.32$\pm$0.14 & 8.38$\pm$0.14 & 8.31$ \pm$0.14 & 8.33$\pm$0.14 & 8.34$\pm$0.14 \\
$O_{32}$ & 1.56$\pm$0.09 & 2.00$\pm$0.12 & 2.61$\pm$0.26 &  2.98$\pm$0.20 & 2.73$\pm$0.17 \\
$O_{32}$$^a$ & 1.28$\pm$0.07 & 1.70$\pm$0.10 & 1.84$\pm$0.18  & 2.28$\pm$0.15 & 2.01$\pm$0.12 \\
$\mathrm{F}_{[\mathrm{SII}]\lambda6717}/\mathrm{F}_{[\mathrm{SII}] \lambda6734}$ & - & 0.7$\pm$0.2 & 0.8$\pm$0.1 & 1.0$\pm$0.1  & 0.7$\pm $0.1 \\
${\mathrm{L}_{\mathrm{H}\alpha}}^c$ & 17 & 17 & 7 & 7 & 7 \\
${\mathrm{L}_{\mathrm{H}\alpha}}^{a,c}$ & 26 & 26 & 15 & 15 & 15 \\
${\mathrm{L}_{\mathrm{H}\alpha}}^{b,c}$ & 66 & 66 & 17 & 17 & 17 \\
${\mathrm{SFR}_{\mathrm{H}\alpha}}^{d}$ & 73 & 73 & 32 & 32 & 32 \\
${\mathrm{SFR}_{\mathrm{H}\alpha}}^{a,d}$ & 113 & 113 & 64 & 64 & 64 \\
${\mathrm{SFR}_{\mathrm{H}\alpha}}^{b,d}$ & 289 & 289 & 75 & 75 & 75 \\
$E(B-V)_{{g-r}}$ & 0.16 & 0.13 & 0.28$^e$ & 0.21$^e$ &  0.24$^{e,f}$ \\
$E(B-V)_{\mathrm{H}\alpha/\mathrm{H}\gamma}$ & 0.45 & 0.45 & 0.28 &  0.28 & 0.28 \\
\enddata
\tablenotetext{a}{These values are corrected for reddening by the  
broadband $g-r$ derived $E(B-V)$ values}
\tablenotetext{b}{These values are corrected for reddening by $E(B-V)$  
values derived from H$\alpha$/H$\gamma$ values and assuming Case B  
recombination}
\tablenotetext{c}{H$\alpha$ luminosity in units of $10^{42}$ ergs $ 
\mathrm{s}^{-1}$, corrected for gravitational lensing, and summed  
across both apertures}
\tablenotetext{d}{SFR in units of $\mathrm{M}_{\sun}$ $ 
\mathrm{yr}^{-1}$, corrected for gravitational lensing and summed  
across both apertures}
\tablenotetext{e}{$g$ and $r$ magnitudes taken from \citet{lin2008}}
\tablenotetext{f}{$g$ and $r$ magnitudes for the full aperture of the  
Clone are taken from the flux of the individual apertures}
\end{deluxetable*}

\subsection{Star-Formation Rate}\label{sec:sfr}

SFRs were calculated from H$\alpha$ luminosities based on the calibrations of 
\citet{kennicutt1998}, but including a normalization factor of 1.8 to convert 
from a \citet{salpeter1955} initial mass function (used by Kennicutt) to a 
\citet{chabrier2003} initial mass function.  Lensing serves to increase the 
measured H$\alpha$ luminosity, and thus the SFR, by the magnification factor of the 
lensed image. In order to correct for lensing, we used the 
published lensing solutions for the Cosmic Horseshoe 
\citep{dye2008,belokurov2007} and the Clone \citep{lin2008}. The published 
magnification factors apply to the entirety of the Einstein ring and not to the 
individual knots that we targeted on our NIRSPEC apertures. We therefore used 
the VLT $R$ and SDSS $r$-band images to find the fraction of the flux in the 
total Einstein ring that fell onto the slit (0.11 for the Cosmic Horseshoe
and 0.30 for the Clone). Under the assumption that the 
rest-frame UV flux in these bands is proportional to the flux in H$\alpha$, we 
divided the measured H$\alpha$ luminosity (for both apertures) by the fraction 
of flux that fell into the slit versus the total flux in the ring. This 
calculation yielded the H$\alpha$ luminosity for the entirety of the ring, which 
we could then convert to a lensed SFR and divide by the magnification to find 
the unlensed SFR. For the Cosmic Horseshoe, the magnification factor for the 
lens is $24\pm2$ \citep{dye2008}, and adopting this lensing model, we calculated 
an unlensed SFR of 73 $M_{\sun}$ yr$^{-1}$ without correcting for reddening,
113 $M_{\sun}$ yr$^{-1}$ with the broadband $g-r$ color reddening correction and 
289 $M_{\sun}$ yr$^{-1}$ for the H$\alpha$/H$\gamma$ reddening correction. For 
the Clone, the magnification factor given for the lens is $27\pm 1$ 
\citep{lin2008}, and under this assumption, the unlensed SFR was 32 $M_{\sun}$ 
yr$^{-1}$ without correcting for reddening, 64 $M_{\sun}$ yr$^{-1}$ with the 
broadband $g-r$ color reddening correction and 75 $M_{\sun}$ yr$^{-1}$ with the 
H$\alpha$/H$\gamma$ reddening correction. Based on the quoted uncertainties
in the magnification factors and $E(B-V)$ values, the formal errors on the lensing-
and dust-corrected SFRs are $\sim 15$\%, yet these
errors do not fully reflect hard-to-quantify uncertainties in 
the fraction of the H$\alpha$ flux that fell into our
slit and the underlying stellar absorption affecting the Balmer lines,
and systematics in the lensing models.

SFRs can also be estimated from rest-frame UV luminosities, 
according to the relation presented in \citet{kennicutt1998}. We use the optical 
photometry from \citet{belokurov2007} and \citet{lin2008} to obtain such 
estimates for the Cosmic Horseshoe and the Clone, respectively, again converting 
from a Salpeter to Chabrier IMF. We calculate an extinction-corrected $SFR_{UV}= 
65-80\: M_{\sun}$ yr$^{-1}$ for the Cosmic Horseshoe, where the range reflects 
the uncertainty in inferred $E(B-V)$ values ($\sim0.13
- 0.16$, see Table \ref{tab:phy}) and magnification factor.
The UV-derived SFR is significantly lower
than the SFR based on H$\alpha$ if the
H$\alpha$ extinction correction is calculated from H$\alpha$/H$\gamma$. Much
better agreement is found when the $E(B-V)$ value used to correct 
H$\alpha$ is based on the $g-r$ color.  We therefore conclude that,
in the case of the Cosmic Horseshoe, the $g-r$ color
is a more reliable tracer of the extinction of both
stars and gas than the H$\alpha$/H$\gamma$ ratio.
For the Clone, we find an 
extinction-corrected $SFR_{UV}= 60-110\: M_{\sun}$ yr$^{-1}$, corresponding to 
$E(B-V)$ values of $\sim0.21 - 0.28$. This UV-derived SFR agrees
very well with the extinction-corrected H$\alpha$ SFR
(regardless of whether $g-r$ or H$\alpha$/H$\gamma$
is used to extinction-correct H$\alpha$).
In conclusion, we present the extinction properties of
the Cosmic Horseshoe and the Clone, along with a comparison
of their H$\alpha$ and UV SFRs. These results 
support the findings of \citet{erb2006c}, that $E(B-V)_{\rm{star}}\simeq E(B-V)_{\rm{gas}}$
in $z\sim 2$ star-forming galaxies.

\subsection{Metallicity}

Our NIRSPEC measurements include a large number of rest-frame optical emission 
line fluxes, enabling the calculation of gas-phase metallicities for our 
objects. The gas-phase metallicity reflects the integrated products of previous 
star formation, modulated by gas inflow and outflow. Typically, oxygen is used 
to probe gas metallicity in star-forming galaxies as it is the most 
abundant heavy element, and is promptly
released into the interstellar medium following a burst of star
formation. Furthermore, the emission lines from the various 
ionization states of oxygen are strong and easily measurable in the rest-frame 
optical. In this paper, solar abundance is defined as 12 + log(O/H) = 8.66 
\citep{allende2002,asplund2004}.

For local galaxies, the oxygen abundance can be inferred from the electron 
temperature by comparing auroral lines (such as [\ion{O}{3}]$\lambda$4363) to 
nebular emission lines (e.g. [\ion{O}{3}]$\lambda$$\lambda$4959,5007). Weak 
auroral lines are difficult to measure in high-redshift galaxies, where the 
ratios of various strong emission lines are used as a proxy for metallicity. 
These ratios have been calibrated as oxygen abundance indicators in local 
\ion{H}{2} regions. The magnification from the strong-lensing has made it 
possible to obtain a large set of high quality measurements of emission lines in 
relatively short exposure times. We therefore can compare three strong-line 
ratios as oxygen abundance indicators: $R_{23}$ $\equiv$ 
([\ion{O}{2}]$\lambda$3727 + [\ion{O}{3}]$\lambda$$\lambda$4959,5007/H$\beta$), 
$N2$ $\equiv$ log([\ion{N}{2}]/H$\alpha$), and $O3N2$ $\equiv$ 
log\{([\ion{O}{3}]$\lambda$5007/H$\beta$) / 
([\ion{N}{2}]$\lambda$6584/H$\alpha$)\}. $N2$ and $O3N2$ have been calibrated using 
direct O/H measurements for local \ion{H}{2} regions by Pettini \& Pagel (2004), 
and we used the $R_{23}$ calibration from \citet{tremonti2004}, which has been 
calibrated with photoionization models. The observed strong-line ratios and 
inferred metallicities are listed in Table \ref{tab:phy}.

The $R_{23}$ indicator was introduced by \citet{pagel1979}, and is widely used 
for measuring local metal abundances if the fluxes of [\ion{O}{3}] and 
[\ion{O}{2}] are known. \citet{tremonti2004} provided an analytical fit to the 
$R_{23}$ - metallicity relation for a set of local SDSS galaxies:

\begin{equation}
12 + \mathrm{log(O/H)} = 9.185 - 0.313x - 0.264x^2 - 0.321x^3
\end{equation}

\noindent where $x = \mathrm{log}\,R_{23}$. Since $R_{23}$ is double valued for 
the larger values of the ratio, this formula is only valid for the high 
metallicity branch of the $R_{23}$-abundance relation. This relationship has a 
1$\sigma$ scatter of 0.05 dex for the sample of star-forming galaxies presented 
in \citet{tremonti2004}, though this error may underestimate the true systematic 
uncertainty \citep{kennicutt2003}. For $R_{23}$, we used the values for the 
H$\beta$ flux that we inferred from the broadband color-derived value of 
$E(B-V)$. $R_{23}$ is sensitive to reddening because of the large wavelength 
difference between [\ion{O}{3}]$\lambda$5007 and [\ion{O}{2}]$\lambda$3727. The 
observed [and extinction corrected] $R_{23}$ values that we calculated for our 
objects were 3.82[4.07], 3.79[3.96] for the two apertures of the Cosmic 
Horseshoe, and 6.86[7.41], 5.25[5.52], and 5.93[6.32] for the two apertures and 
the full aperture of the Clone. The $R_{23}$ values for apertures 1 and 2 of the 
Cosmic Horseshoe correspond to metallicities of 12+log(O/H) = 8.85[8.82] and 
8.85[8.84], respectively. These are high relative to solar metallicity. For the 
Clone, the inferred metallicities for apertures 1, 2 and the full aperture are 
12+log(O/H) = 8.55[8.50], 8.70[8.68], and 8.64[8.60], respectively. These values 
range from slightly subsolar to solar.

The $N2$ indicator is related to the oxygen abundance by:

\begin{equation}
12 + \mathrm{log(O/H)} = 8.90 + 0.57 \times \mathrm{N}2
\end{equation}

\noindent which is valid for 7.50 $<$ 12 + log(O/H) $<$ 8.75. 
The relationship has a 1$\sigma$ scatter of $\pm$0.18 dex on the 
y-intercept \citep{pettini2004}. We found values of $N2 = -1.04$ 
and $-0.79$ (12+log(O/H) = 8.31 and 8.45) for the Cosmic Horseshoe, 
and $-0.75$, $-0.79$, and $-0.72$ (12+log(O/H) = 8.47, 8.45, and 8.50) 
for the Clone. These values all indicate a subsolar gas metallicity.

We also measured values of the $O3N2$ indicator for our objects. 
This indicator is related to oxygen abundance by:

\begin{equation}
12 + \mathrm{log(O/H)} = 8.73 - 0.32 \times \mathrm{O}3\mathrm{N}2
\end{equation}

\noindent which is valid for 8.12 $<$ 12+log(O/H) $<$ 9.05, with a 1$\sigma$ 
scatter of $\pm$0.14 dex \citep{pettini2004}. The errors shown for our oxygen 
abundances are dominated by the systematic uncertainties from the calibration, 
as opposed to measurement errors. For $O3N2$, we used the values of H$\beta$ that 
were calculated from the $g-r$ color, as described in \S \ref{sec:dust}. The 
values that we calculated are 1.28, 1.08 (12+log(O/H) = 8.32 and 8.38) for the 
Cosmic Horseshoe, and 1.31, 1.26, and 1.23 (12+log(O/H) = 8.31, 8.33, and 8.34) 
for the Clone. These values indicate subsolar metallicities in the gas. The $N2$
and $O3N2$ indicators do not need to be dereddened due to the close spacing of the 
[\ion{N}{2}], H$\alpha$ lines and [\ion{O}{3}], H$\beta$ lines.

The $N2$ and $O3N2$ indicators yield consistent metallicities of $\sim0.5$ solar for 
both the Cosmic Horseshoe and the Clone. However, the $R_{23}$ indicator points 
to a significantly higher, and supersolar, metallicity in the Cosmic Horseshoe. 
Such systematic discrepancies between $R_{23}$ and other indicators are well 
known among local star-forming galaxies, and reflect the current limitations of 
strong-line abundance indicators \citep{kewley2008, kennicutt2003}. For the 
Clone, $R_{23}$ indicates a metallicity that is only slightly higher than, and 
statistically consistent with, the values implied by $N2$ and $O3N2$. As we will 
discuss in \S \ref{sec:offset}, the differences in metallicity for the Cosmic 
Horseshoe and the Clone based on $R_{23}$ may reflect the relative positions of 
these objects with respect to the emission-line excitation sequence of local 
objects.

\subsection{Ionization Parameter}

The local ionization state in an \ion{H}{2} region is often characterized by the 
ionization parameter, $U$, which is the ratio between the density of ionizing photons 
and the density of hydrogen atoms. Commonly, ionization parameters are estimated 
using $O_{32}$ = ([\ion{O}{3}]$\lambda$$\lambda$4959,5007 / 
[\ion{O}{2}]$\lambda$$\lambda$3726,3729), and corrected for reddening. We 
estimate extinction based on the observed $g-r$ colors. Our values of $O_{32}$, 
both uncorrected [and corrected] for extinction, are 1.56[1.28], 2.00[1.70] for 
the two apertures of the Cosmic Horseshoe, and 2.61[1.84], 2.98[2.28], and 
2.73[2.01] for the two apertures and the full aperture of the Clone. 
\citet{lilly2003} showed that a vast majority of local objects (from a Nearby 
Field Galaxy Survey $B$-selected local galaxy sample from \citet{jansen2000}) 
have a value of $O_{32} < 1$, while the available data for objects at $z > 2$ 
\citep{pettini2001} indicate $O_{32} > 1$ \citep{pettini2001}. 

The value of $O_{32}$, however, is dependent not only on ionization parameter, 
but also, to a lesser extent, on metallicity \citep{kewley2002,brinchmann2008}.
\citet{kewley2002} provide formulae for relating observed values of 
$O_{32}$ to ionization parameter, $\log(U)$, using photoionization models
spanning a range of metallicities. 
We apply the relations for both $0.5$ and $1.0 Z_{\sun}$ models
to translate our extinction-corrected $O_{32}$ measurements into
$\log(U)$ values. The adopted models reflect the range of metallicities
inferred for the Cosmic Horseshoe and the Clone.
At higher metallicities, a given value of $O_{32}$ corresponds to a
higher ionization parameter, so the $0.5 Z_{\sun}$ models yield a lower
bound to the inferred ionization parameters. Assuming $Z=1.0 Z_{\sun}$,
we find $\log(U)$ ranging from -2.4 to -2.5 and -2.3 to -2.4 for the
apertures of the Cosmic Horseshoe and the Clone, respectively.
At $Z=0.5 Z_{\sun}$, the corresponding ionization parameters range from
$\log(U)=-2.7$ to -2.8 and -2.5 to -2.6, i.e., $0.2-0.3$ dex lower.
The $\log(U)$ values inferred for $0.5$ and $1.0 Z_{\sun}$
models are very similar to the high ionization parameters
(relative to local SDSS galaxies)
inferred by \citet{brinchmann2008} for the $z>2$ objects
in \citet{pettini2001}.

\begin{figure*}
\epsscale{1.} 
\plotone{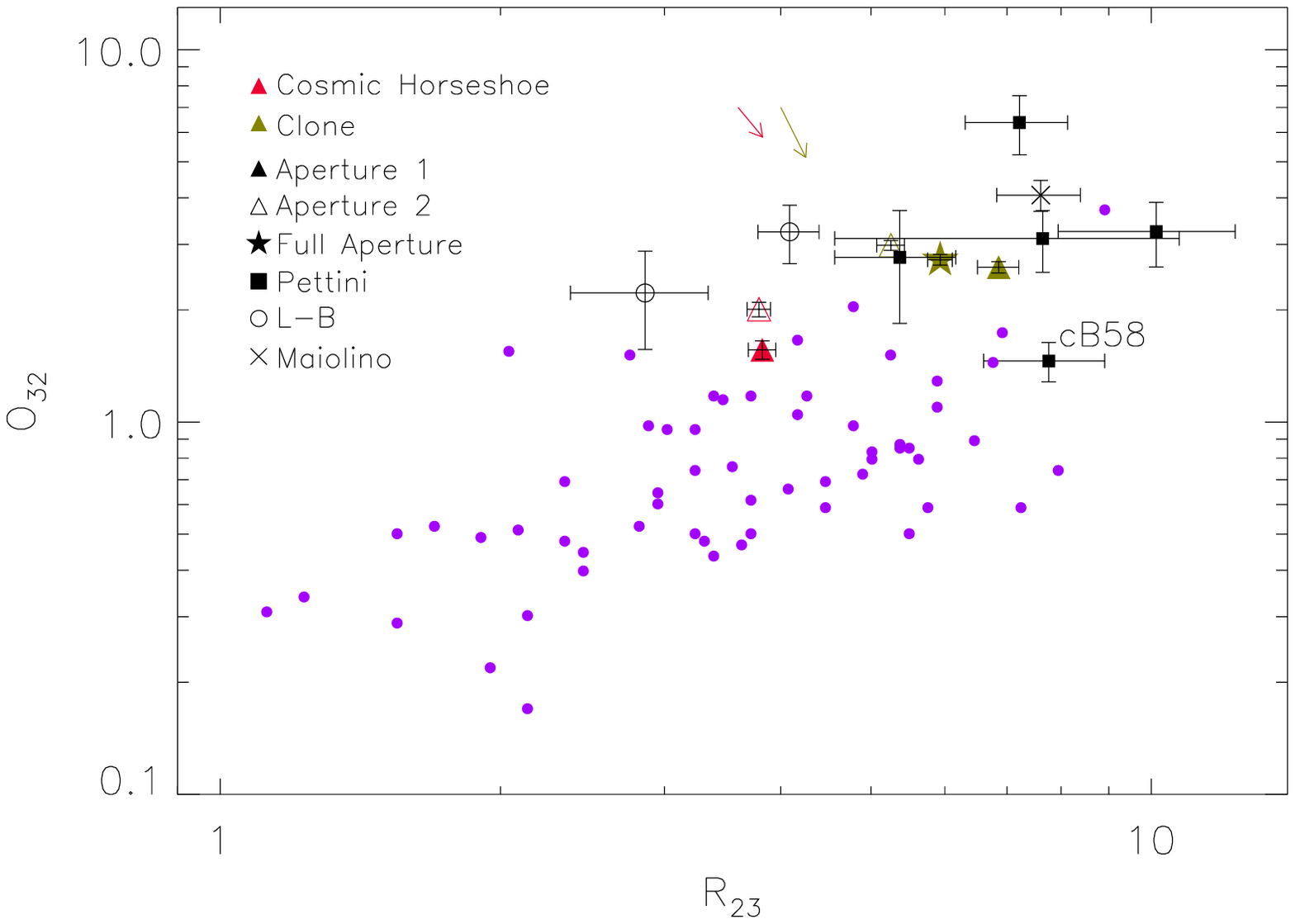} 
\caption{$O_{32}$ versus $R_{23}$ diagram. The apertures of the Cosmic Horseshoe
and the Clone are plotted along with star-forming galaxies at intermediate
and high redshift.  A selection of $0.47 < z < 0.92$ objects 
from \citet{lilly2003} are shown in purple. High-redshift points
include five $z\sim3$ LBGs from \citet{pettini2001}, two lensed $z\sim 1.9$ objects
from \citet{lemoine2003}, and a composite spectrum of nine $z\sim 3.5$ LBGs
from \citet{maiolino2008}. Symbols are described in the legend. The values on
this plot are not corrected for dust extinction, however the red and
green arrows indicate, respectively, the small impact
of extinction corrections on the Cosmic Horseshoe and Clone data points.
Such dust corrections will shift points downwards and to the right.
The objects at $z\geq2$ are systematically displaced on average with respect
to the lower-redshift sample towards
larger values of $O_{32}$, at fixed values of $R_{23}$.
This empirical trend is evidence
for higher ionization parameters at fixed metallicity \citep{kewley2002}.
\label{fig:O32R23}}
\epsscale{1.}
\end{figure*}

\subsection{Electron Temperature and Density}

The [\ion{S}{2}]$\lambda$$\lambda$6717,6732 line ratio is sensitive to the 
electron density of an ionized gas. We used this line ratio in our objects to 
calculate the electron density for the ionized nebular gas. Similarly, the ratio 
([\ion{O}{3}]$\lambda$$\lambda$4959, 5007 / [\ion{O}{3}]$\lambda$4363) is used 
to measure the electron temperature, as [\ion{O}{3}] has an energy-level 
structure that results in upper level emission lines with different excitation 
energies, whose relative strengths thus depend strongly on temperature. We used 
the IRAF procedure TEMDEN to calculate a range of densities and temperatures 
based on these line ratios.

Upper limits on the electron temperatures were measured from the [\ion{O}{3}] 
line ratios, since we were only able to measure upper limits for the 
[\ion{O}{3}]$\lambda$4363 line flux. For the lowest densities ($\sim$100 
cm$^{-3}$), the upper limits on the temperatures were 12,900 K and 12,200 K for 
the two apertures of the Cosmic Horseshoe and 13,200 K, 9,700 K and 10,000 K for 
the two apertures and the full aperture of the Clone, respectively. Changes in 
the density over the range 100 to 5000 cm$^{-3}$ altered the calculated 
temperatures only very slightly.

We calculated densities from the measured [\ion{S}{2}] line ratios in all 
apertures except for aperture 1 of the Cosmic Horseshoe, which had insufficient 
S/N. We also summed both apertures of the Cosmic Horseshoe to
obtain an average, higher S/N, [\ion{S}{2}] line ratio estimate for the galaxy.
In the calculation of densities we assumed a standard temperature of 
10,000 K, consistent with our limits on electron temperatures. For
each object the $\pm 1\sigma$ range in electron density
was then derived from errors on the observed line ratios.
For the Cosmic Horseshoe, Aperture 2,
the $\pm 1\sigma$ range in electron density is $840-6900 \mbox{ cm}^{-3}$;
for the summed Cosmic Horseshoe apertures, the corresponding
range is $320-1600 \mbox{ cm}^{-3}$. For the Clone, apertures
1, 2, and full, we find ranges of $1110-2960\mbox{ cm}^{-3}$,
$530-1020 \mbox{ cm}^{-3}$, and $1270-2540\mbox{ cm}^{-3}$. Despite
the large uncertainties in the precise value of inferred electron density,
the empirical fact remains that
the high-redshift lensed targets exhibit systematically lower
doublet ratios from what is typically observed in emission-line galaxies
in the local universe \citep{liu2008}.

\section{Analysis}\label{sec:analysis}

\subsection{Comparison to the $z\sim2$ Population\label{sec:comparison}}

Recent studies have been undertaken to understand the physical conditions of 
$z\sim2$ objects \citep{erb2006a,liu2008} and the observations presented here 
enable comparisons of the Cosmic Horseshoe and the Clone with the larger 
population of star-forming galaxies at similar redshifts. The Cosmic Horseshoe 
and the Clone, based on their $ugr$ colors, satisfy the criteria for UV-selected 
objects at $z\sim2$ \citep{steidel2004}. In the following analysis, we compare 
the physical quantities calculated from the rest frame optical line fluxes to 
those calculated in other high-redshift objects.

SFRs in high redshift objects have been measured in a variety of ways. We 
calculated lensing-corrected values for the SFR of $\sim 113 \: M_{\sun}$ 
yr$^{-1}$ based on our method of dereddening the H$\alpha$ fluxes for the Cosmic 
Horseshoe (i.e., from the $g-r$ color) and $64-75 \: M_{\sun}$ yr$^{-1}$ based on our method of 
dereddening for the Clone (i.e., from both the $g-r$ color and H$\alpha$/H$\gamma$
ratio). The average SFRs for a population of $z\sim2$ 
galaxies from \citet{erb2006c} is $\langle SFR_{H\alpha} \rangle = 31 \pm 18$ 
$M_{\sun}$ yr$^{-1}$. Our lensed targets are therefore more actively star-forming
than average, but within the range spanned by the sample in \citet{erb2006c}. 
However, we caution that our calculated 
values depend on the magnification given by the particular lensing models that 
we used. For comparison, based on the H$\alpha$ flux and magnification factor of 
$\sim30$ \citep{teplitz2000, seitz1998} and with the assumption of a Chabrier 
initial mass function and \citet{calzetti2000} dust extinction law with an 
estimated $E(B-V)_{\rm{gas}} = 0.06$ \citep{siana2008}, the SFR for MS1512-cB58 is 
$SFR(H\alpha) = 14$ $M_{\sun}$ yr$^{-1}$.

\citet{erb2006b} measure an average velocity dispersion for a sample of $z\sim2$ 
objects (with AGN removed) of $\langle \sigma \rangle = 108 \pm 5$ km s$^{-1}$, 
and a standard deviation of 86 km s$^{-1}$. Our objects show a smaller velocity 
dispersion than the average, but are within the sample's large standard 
deviation. While SDSS J0901+1814, in particular aperture 2, displays AGN-like 
line ratios, the velocity dispersions are very close to the average values for 
other $z\sim2$ objects, indicating that SDSS J0901+1814 could be a narrow-line 
AGN. The dust extinction values that we infer from the $g-r$ color
for the Cosmic Horseshoe are similar to
the mean $E(B-V)$ value of $0.16 \pm 0.01$ from \citet{erb2006b},
while the Clone appears to be dustier than average, based on
its $g-r$ color and H$\alpha$/H$\gamma$ ratio. Both objects are well within
the range of extinction values spanned by $z\sim 2$ star-forming galaxies.

Our values for $N2$ ranged from -0.72 to -1.04, within the interval spanned by 
the sample of $z\geq2$ galaxies from \citet{erb2006a}. Values of $R_{23}$ span 
from 4.0 to 7.4, resulting in a range of metallicities that are slightly 
subsolar to slightly greater than solar, between 12+log(O/H) = 8.5 to 8.8. 
Values for $R_{23}$ from \citet{pettini2001} for LBGs at $z\sim3$ are within 
this range, between 4.0 and 12.3, including a value of $R_{23} = 8.3$ for MS 
1512-cB58. \citet{lilly2003} plot star-forming galaxies at intermediate redshift 
($0.47 < z < 0.92$) by their values for $O_{32}$ versus $R_{23}$, and we 
reproduce their diagram in Figure \ref{fig:O32R23}, adding the Cosmic Horseshoe and
Clone. To place our new observations in the context of other, similar
measurements at high redshift, we also plot five LBGs at $z\sim3$ from \citet{pettini2001},
two lensed objects at $z\sim 1.9$ from \citet{lemoine2003},
and the composite $z\sim 3.5$ spectrum from \citet{maiolino2008},
based on nine individual LBGs. In this 
plot, note that all of the values are as measured, and none have been 
dereddened. The effect of correcting the points for 
the Cosmic Horseshoe and Clone for dust extinction
is indicated with reddening vectors. This diagram shows how objects at $z\geq 2$ have 
larger values of $O_{32}$ (indicative of high values of the ionization 
parameter) for a given value of R$_{23}$ (and thus, metallicity),
relative to the sample from \citet{lilly2003}.

A common property of $z\geq 2 $ star-forming galaxies
is the kinematic evidence for star-formation feedback
in the form of large-scale gas outflows \citep{pettini2001,shapley2003}.
With measurements of the nebular, systemic redshifts for both the 
Cosmic Horseshoe and Clone, as well as rest-frame UV absorption-line
spectra, it is possible to search for the signature
of outflows. Using Keck II Echelle Spectrograph and
Imager \citep[ESI;][]{sheinis2002} rest-frame ultraviolet spectra of the Cosmic
Horseshoe and the Clone, \citet{quider2009} and Quider et al. (in preparation)
analyze several of the strongest interstellar
absorption features for kinematic evidence of large-scale gas outflow.
In this analysis, redshifts measured for interstellar absorption
lines were compared with those of H$\alpha$ and UV stellar photospheric features.
For both targets, the slits were placed in the same position that was used for 
the NIRSPEC observations, and the spectra from the two apertures were summed to 
increase the signal-to-noise ratio. In the Cosmic Horseshoe,
the average outflow velocity for the low-ionization
interstellar absorption lines (e.g., \ion{Si}{2} $\lambda$1260, 
the \ion{O}{1} + \ion{Si}{2} $\lambda$1303 doublet,
\ion{Si}{2} $\lambda$1527, \ion{Al}{2} $\lambda$1671)
is $\langle v_{out,low} \rangle = 146$ km s$^{-1}$,
and  $\langle v_{out,high} \rangle = 167$ km s$^{-1}$ for
the high-ionization features (e.g., \ion{Si}{4} $\lambda\lambda $1394,1403,
\ion{Al}{3} $\lambda$1855, \ion{Al}{3}
$\lambda$1863). In the Clone, three low-ionization iron
lines (\ion{Fe}{2} $\lambda$2344, 2383, and 2600) were used
to measure interstellar kinematics.  These three lines indicate similar
offsets, and the average outflow velocity is $\langle v_{out,low} \rangle = 154$
km s$^{-1}$. The outflow velocities observed in the Cosmic Horseshoe
and Clone are consistent with those calculated for star-forming
galaxies at $z\sim2$ \citep{steidel2004}. Also, for comparison, the outflow
velocity observed in MS 1512-cB58 is $\sim255$ km s$^{-1}$ \citep{pettini2002}.

The H$\alpha$ velocity dispersion, $\sigma$, allows us to calculate the virial 
masses of our objects when combined with estimates of their sizes, using the 
half-light radius, $r_{1/2}$. We use the equation given by \citet{pettini2001} 
for virial mass, assuming an idealized case of a sphere of uniform density:

\begin{equation}
M_{\mathrm{vir}} = 1.2 \times 10^{10} M_{\sun} \frac{\sigma^2}{(100 \,\mathrm{km}\,\mathrm{s}^{-1})^2} \frac{r_{1/2}}{\mathrm{kpc}}
\end{equation}

\noindent The lensing models for both the Cosmic Horseshoe and the Clone yield 
estimates of the intrinsic half-light radii, allowing us to calculate the 
virial masses of the objects. 

The lensing model for the Cosmic Horseshoe \citep{dye2008} produces a source 
plane image with two objects, a ``northern" and a ``southern" source. These 
objects are lensed into ring-like distributions in the image plane, with 
slightly different morphologies. Accordingly, the relative contribution of the 
two sources varies as a function of position in the combined Einstein ring. The 
fainter, northern source, according to the \citet{dye2008} model, contributes 
mostly to the portion of the Einstein ring that is not probed by our slit 
position, although the light from this source does contribute a small fraction 
of the emission in what we call aperture 1. The southern source, the more 
prominent of the two, is the only contributor to the emission in aperture 2 and 
constitutes the majority of the light in aperture 1. From the composite of the 
reconstructed source plane image, we estimated a half-light radius of the 
southern source of $\sim0\secpoint3$, which, at this redshift, and, for our 
adopted cosmology, corresponds to a size of $r_{1/2, CH} = 2.5$ kpc. Thus, using 
the velocity dispersion for the second aperture, since it corresponds only to 
the southern source in the reconstructed image, we calculate a virial mass of 
the Cosmic Horseshoe of $M_{vir,CH} = 1.0 \times 10^{10} M_{\sun}$.

For the Clone, the lensing models described in \citet{lin2008} and Allam et al. 
(2009, in preparation) yield estimates of the half-light radius in the 
reconstructed source plane of $\sim0\secpoint3$. At this redshift, the 
corresponding physical radius is $r_{1/2, Clone} = 2.9$ kpc. The virial mass we 
calculated for the Clone, using the velocity dispersion from the full aperture, 
was $M_{vir,Clone} = 2.2 \times 10^{10} M_{\sun}$. Both the dynamical masses 
calculated for the Cosmic Horseshoe and the Clone are similar to the dynamical 
masses for objects at $z\sim2$ from \citet{erb2006b}.

While the line flux ratios for the Clone are very similar between the apertures, 
the \citet{dye2008} lensing model might help to explain the difference in the 
line flux ratios observed between the two apertures of the Cosmic Horseshoe. For 
instance, as seen in Table \ref{tab:emi} and Figure \ref{fig:specCH}, the 
[\ion{N}{2}] and H$\alpha$ lines vary in strength with relation to each other. 
One possible hypothesis concerns the level of contribution from the ``northern" 
source to the first aperture of the lensed ring. If the gas in the fainter, 
northern source was at a lower metallicity, it would slightly lower the 
[\ion{N}{2}]/H$\alpha$ line ratio of aperture 1 compared to aperture 2, which is 
what is observed in our data. Future analysis will be undertaken with the 
lensing model in order to disentangle the contribution from each object in the 
source plane.

In summary, we have shown that the SFRs, extinction values, metallicities, 
outflow velocities, and dynamical masses for our objects are relatively typical 
of those found among UV-selected star-forming galaxies at $z\sim2$ 
\citep{steidel2004}. We conclude that the Cosmic Horseshoe and the Clone are a 
representative sample of the luminous segment of the population of $z\sim2$ objects. Future 
multi-wavelength observations of the broadband spectral energy distributions of 
the lensed targets will allow us to characterize their stellar masses and ages 
and therefore evolutionary states.

\subsection{The Offset in the [\ion{N}{2}]/H$\alpha$ vs. [\ion{O}{3}]/H$\beta$ diagram 
for High-Redshift Galaxies}\label{sec:offset}

\begin{figure*}
\epsscale{1.} 
\plotone{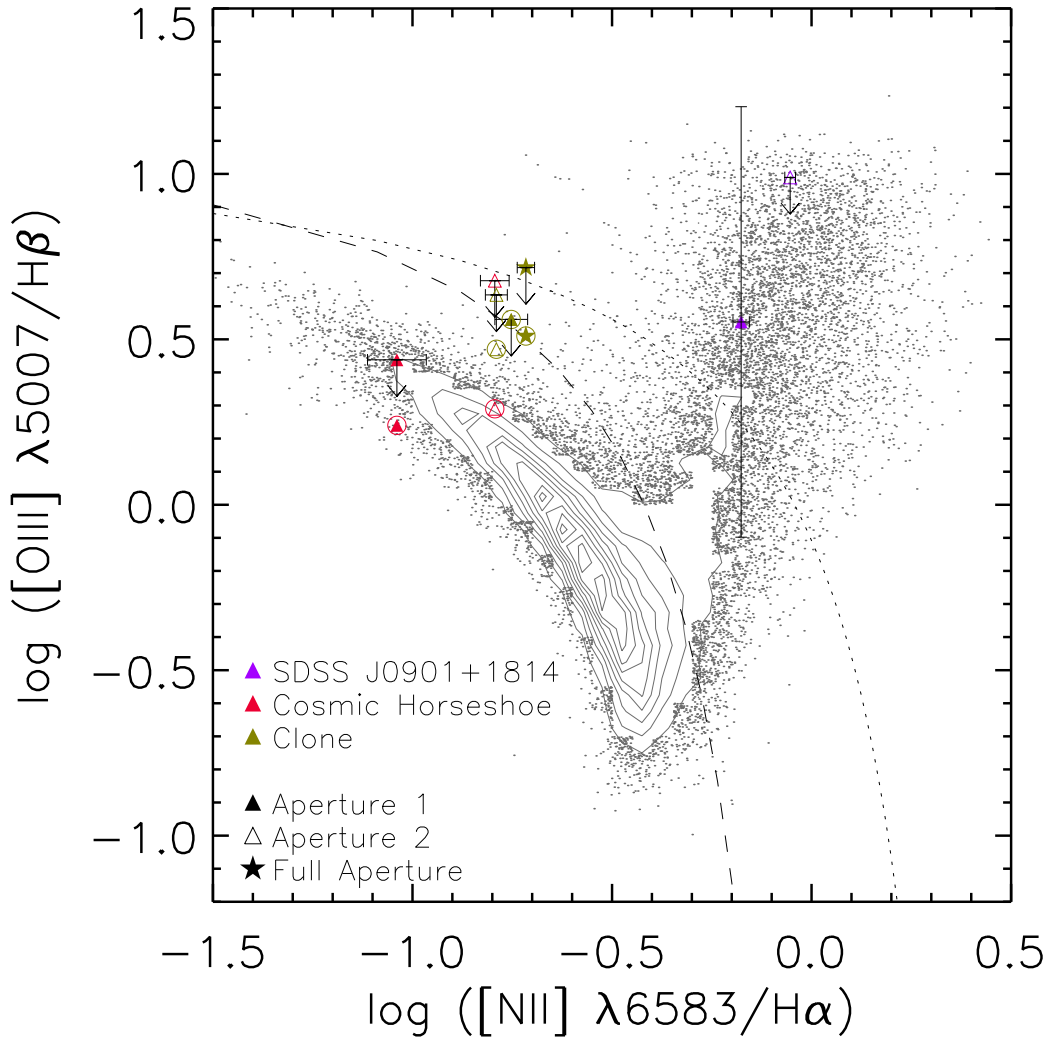} 
\caption{\ion{H}{2} region diagnostic plot of log[\ion{N}{2}]/H$\alpha$ and 
log[\ion{O}{3}]/H$\beta$ emission line ratios. The grey points and contours 
represent SDSS local AGN and star forming galaxies that satisfy the criteria 
described in \S \ref{sec:offset}, while the lensed objects are as labelled in
the legend. The 
circled objects are those where the H$\beta$ flux values were inferred from the 
$g-r$ color and H$\alpha$ fluxes. SDSS J0901+1814 points show large error bars 
due to the presence of a sky line on the H$\beta$ line in both apertures. The 
dashed line is an emprical demarcation between star-forming regions and AGN from 
\citet{kauffmann2003b} based on the SDSS galaxies, whereas the dotted line is 
the theoretical limit for star-forming galaxies from \citet{kewley2001a}. Local 
star-forming galaxies and \ion{H}{2} regions exist in a well-defined excitation 
sequence due to photoionization by massive stars, and these are found to the 
left and below the curves. The apertures of the Clone, even with the inferred 
values of H$\beta$, are on average offset from this excitation sequence.
\label{fig:BPT}} 
\epsscale{1.}
\end{figure*}

Recently, \citet{shapley2005}, \citet{erb2006a}, and \citet{liu2008} have 
presented evidence indicating a difference between \ion{H}{2} regions in 
high-redshift galaxies and those in local galaxies. The physical conditions in 
these regions are often probed by using the well known empirical diagnostic 
diagram of Baldwin, Phillips, \& Terlevich (1981) (hereafter BPT). This diagram 
separates star-forming galaxies and AGN according to the optical line ratios 
[\ion{N}{2}]/H$\alpha$ and [\ion{O}{3}]/H$\beta$; star-forming galaxies fall in 
a region of lower [\ion{N}{2}]/H$\alpha$ for a range of values of 
[\ion{O}{3}]/H$\beta$. It has been demonstrated that a fraction of high redshift 
star-forming galaxies at $z \sim 1 - 2$ lie offset from the local population of 
\ion{H}{2} regions and star-forming galaxies, displaced towards higher 
[\ion{N}{2}]/H$\alpha$ and [\ion{O}{3}]/H$\beta$ values. The strongly lensed, 
$z\sim2$ objects are plotted on the BPT diagram in Figure \ref{fig:BPT} along 
with local emission-line objects from the fourth data release (DR4) of the SDSS 
\citep{york2000,adelman2006}. The SDSS objects were selected to have 
[\ion{N}{2}], H$\alpha$, [\ion{O}{3}], and H$\beta$ line detections with S/N $>$ 
5. Also plotted on the diagram are curves designed to separate star forming 
galaxies and AGN; one line is from \citet{kauffmann2003b} that demarcates 
star-forming galaxies and AGN on an empirical basis, and the other from 
\citet{kewley2001a} that represents a limit to the line-flux ratios that can be 
produced for star forming regions from photoionization plus stellar population 
synthesis models.

For both the Cosmic Horseshoe and the Clone, we plot two points
for each aperture, one for each of 
method of calculating H$\beta$ fluxes. First, we plot upper limits on 
[\ion{O}{3}]/H$\beta$ to reflect the fact that the measured H$\beta$ is a lower 
limit due to sky-subtraction residuals. Next, we plot circled symbols 
using the values of H$\beta$ flux inferred 
from H$\alpha$ and the $g-r$ color as described in \S \ref{sec:dust}. While SDSS 
J0901+1814 lies in the region occupied by local AGN, the two apertures of the 
Clone both appear in the area above the local star-forming regions. For the 
inferred H$\beta$ values, the line ratios for the Clone are shifted towards 
higher [\ion{O}{3}]/H$\beta$ by $0.2 - 0.3$ dex, relative to star-forming SDSS 
galaxies with similar [\ion{N}{2}]/H$\alpha$ ratios. Alternatively, the Clone is 
shifted towards higher [\ion{N}{2}]/H$\alpha$ by 0.6 - 0.7 dex compared to SDSS 
galaxies with similar [\ion{O}{3}] / H$\beta$ ratios. 
The interpretation is less straightforward for the Cosmic Horseshoe.
Indeed, the [\ion{O}{3}]/H$\beta$ upper limit for its second aperture is also offset
from the local emission-line sequence, yet adopting the H$\beta$ value based
on the H$\alpha$ flux and $g-r$ color places this aperture on the top edge of
the locus of low-redshift galaxies. On the other hand, 
using the H$\alpha$/H$\gamma$ ratio to estimate the H$\beta$ flux results
in an [\ion{O}{3}]/H$\beta$ ratio that is midway between the two
plotted symbols for this aperture, and, again, offset with 
respect to the local emission-line sequence. 
Our best estimate for aperture 2 of the Cosmic Horseshoe is that
its point on the BPT diagram lies somewhere between top of the local emission-line
sequence and the symbol indicating the upper limit for the aperture.
In contrast, the point for aperture 1 of the Cosmic Horseshoe
is consistent, or even below, the local emission-line sequence, depending on which
value of H$\beta$ is adopted. Identifying the origin of the emission-line
ratio variations within the Cosmic Horseshoe clearly requires a more detailed
investigation, in concert with the lensing model for this system. Such
analysis is outside the scope of the current work.

We have not corrected our 
H$\beta$ fluxes for possible stellar absorption at this wavelength. Stellar 
absorption would lead to an underestimate of the flux, and push the points 
upward on the BPT diagram. However, as noted in \citet{shapley2005} and 
\citet{erb2006a}, the stellar H$\beta$ absorption line should have an equivalent 
width of only $W_{abs} \leq 5$ \AA \citep{charlot2002}. Given that the Clone has
$W_{H\beta} \geq 20-30$ \AA, and the Cosmic Horseshoe does not have significant continuum
detected in the spectral region near H$\beta$,
stellar absorption is not significant. Furthermore, we defer application
of this correction until the underlying stellar populations
for the Cosmic Horseshoe and the Clone have been better constrained
by deeper spectroscopy and multi-wavelength photometry.

While the Cosmic Horseshoe does not display the same clear offset as the Clone on the 
BPT diagram, when considered as a {\it population}, the sample of $z\sim 2$ 
objects with line ratio measurements \citep[i.e., those presented in this paper 
as well as those in][]{erb2006a} appears to be offset on average from local 
galaxies. We have demonstrated that the Cosmic Horseshoe and Clone are typical 
of $z\sim 2$ star-forming galaxies in many respects, and also
display several striking differences with respect to local galaxies
in terms of their physical properties. As such, we will use the 
special insights afforded by gravitational lensing into the physical conditions 
in the Cosmic Horseshoe and Clone to explain the properties of the $z\sim 2$ 
population as a whole. We recognize the simplification associated
with drawing conclusions based on only two objects, and the
fact that a significantly larger sample of galaxies with the complete
set of rest-frame optical emission lines is required to characterize
the full $z\sim 2$ population. The discussion of the Cosmic
Horseshoe and the Clone that follows represents the initial
step in the direction of such a statistical study.

An important point of contrast between local star-forming galaxies in the SDSS 
and our high-redshift targets is the rate of star-formation activity. 
\citet{shapley2005} report an aperture-corrected mean SFR for a local sample of 
SDSS galaxies of 2.0 $M_{\sun}$ yr$^{-1}$. These authors also noted that offset 
objects in the BPT diagram (both at high redshift and in the local SDSS sample) 
seem to exhibit elevated SFRs relative to the local average, and this is further 
demonstrated in a small sample of extreme SDSS galaxies in \citet{liu2008}. We 
estimate lensing-corrected SFRs in the Clone and Cosmic Horseshoe that are even 
higher than those presented in \citet{shapley2005}, more than an order of 
magnitude larger than the average SDSS SFR.

The higher rate of star formation might lead to a larger reservoir of ionizing 
photons, which is reflected in the ionization parameter. \citet{brinchmann2008} 
demonstrated that the diagnostic line ratios in the BPT diagram are strongly 
dependent on the ionization parameter. However, while previous discussion of the 
offset has only provided possible explanations that could lead to this effect, 
our measurement of a host of lines, such as 
[\ion{O}{2}]$\lambda$$\lambda$3726,3729 in the magnified objects, allows us to 
test this discussion more quantitatively. To date, only a small sample of 
ionization parameter measurements have been obtained for objects at $z\geq2$ 
\citep{pettini2001, lemoine2003,maiolino2008}. In both the Cosmic Horseshoe and the Clone, our 
measurements of $O_{32}$ indicate higher values of the ionization parameter than 
seen in even the most extreme low-$z$ UV selected galaxies from 
\citet{contini2002}, or a sample of star-forming galaxies at intermediate redshift 
($0.47 < z < 0.92$) from \citet{lilly2003}. Figure \ref{fig:O32R23}
illustrates this point very clearly, in that, at fixed metallicity (i.e., $R_{23}$)
high-redshift galaxies are offset on average towards significantly higher
$O_{32}$ (i.e., ionization parameter). If $O_{32}$ values were considered in isolation, the differences
in $O_{32}$ between low- and high-redshift samples could simply be attributed
to metallicity differences. However, the addition of $R_{23}$ measurements indicates
that there is a real effect towards systematically higher ionization parameters
at high redshift.

Another idea discussed in \citet{brinchmann2008} and \citet{liu2008} is that the 
higher ionization parameter that might lead to this offset on the BPT diagram is 
due to a higher density in the \ion{H}{2} regions. The 
[\ion{S}{2}]$\lambda$$\lambda$6717,6732 line ratio allowed us to measure the 
density of the star forming regions in the Cosmic Horseshoe and the Clone, and 
the low resulting ratios indicated high densities of about $\sim10^3$ cm$^{-3}$, 
which is an order of magnitude higher than the values encountered in local 
starbursts \citep{kewley2001b}. The ESI spectra analyzed by \citet{quider2009}
and Quider et al. (in preparation) also show another independent 
measure of the density in the form of the 
[\ion{C}{3}]$\lambda$1907/\ion{C}{3}]$\lambda$1909 line ratio. With this line 
ratio, values of about 1.5 and above are in the low density regime, while values 
from 0.2 and lower are in the high density regime. For the Cosmic Horseshoe, 
deblending the two lines in a smoothed spectrum yields a line ratio of $1.1\pm 0.2$, and 
for the Clone, the line ratio is $1.2\pm 0.2$. The values for the Cosmic Horseshoe and 
the Clone indicate densities ranging from $5000-22000 \mbox{ cm}^{-3}$ 
and $3000-17000\mbox{ cm}^{-3}$, respectively. These are at least as high as those derived from the 
[\ion{S}{2}] doublet \citep{agn2}. \citet{brinchmann2008} calculate that 
densities of this order would account for an increase in the ionization 
parameter, which might lead to the observed offset on the BPT diagram.

\section{Conclusions}\label{sec:conclusions}
We present NIRSPEC rest-frame optical spectra of three strongly-lensed $z\sim2$ 
galaxies. These include SDSS J0901+1814 ($z = 2.26$), which, due to its observed 
line ratios, is possibly contaminated by an AGN; and two star-forming galaxies, 
the Cosmic Horseshoe ($z = 2.38$) and the Clone ($z = 2.00$). The general 
physical properties of the Cosmic Horseshoe and the Clone are representative of 
the properties found for star-forming galaxies at the same redshift. 
Specifically, we have measured the SFR from the H$\alpha$ luminosity, corrected 
for reddening and magnification from the lensing, and found it to be high yet 
typical of other measurements of high-redshift star formation (SFR = $\sim110$ 
$M_{\sun}$ yr$^{-1}$ for the Cosmic Horseshoe and $\sim70$ $M_{\sun}$ yr$^{-1}$ for the 
Clone). We have also used the $R_{23}$, $N2$, and $O3N2$ methods to 
calculate the metallicity of the \ion{H}{2} regions, and found that, while there 
are differences among the metallicities calculated from these indicators, we can 
still constrain the metalliticies to range from slightly sub-solar to solar. The 
dynamical masses calculated from H$\alpha$ velocity dispersions and the 
half-light radii of the reconstructed sources are on the order of $10^{10}$ 
$M_\sun$, which is typical of the dynamical masses of UV-selected star-forming 
galaxies at $z\sim2$. Finally, ESI rest-frame UV spectra provide evidence for 
the existence of outflowing gas with a velocity on the order of $\sim150-200$ km 
s$^{-1}$ for the Cosmic Horseshoe and the Clone, 
which are also similar to outflows seen in other high-redshift galaxies 
\citep{pettini2001,steidel2004,adelberger2003}. In the future, it will be 
valuable to model the stellar populations of our lensed targets using 
multi-wavelength broadband photometry, and therefore obtain constraints on their 
stellar masses and ages.

The combination of strong lensing and NIRSPEC observations has allowed us to 
probe physical conditions that to date have been largely unexplored at $z\geq2$. 
The measurements of the [\ion{S}{2}]$\lambda$$\lambda$6717,6732 line ratio 
indicated high densities ($\sim10^3$ cm$^{-3}$) in these regions, in agreement 
with the densities derived from 
[\ion{C}{3}]$\lambda$1907/\ion{C}{3}]$\lambda$1909 line ratios for both objects. 
Large ionization parameters were measured by using $O_{32}$, which further 
indicates the high ionization state of the gas in high-redshift objects. The 
high values for the ionization parameter, density, and SFR in both 
the Cosmic Horseshoe and the Clone help in understanding why, on average, the 
population of high-redshift objects are offset on the BPT diagram, a standard 
diagnostic for star-forming regions \citep{shapley2005,erb2006a,liu2008}. The 
exceptional data quality enables the measurement of quantities that previously 
were only speculation for $z\sim2$ galaxies. These results offer more concrete 
evidence of the different conditions under which star formation occurs in 
galaxies at high redshift, yet a statistical sample is still required
to place the results on firmer ground. Future observations with the Multi-Object 
Spectrometer for Infra-Red Exploration (MOSFIRE) instrument planned for the Keck 
I telescope will allow for the assembly of a much larger sample of rest-frame 
optical emission lines for this type of analysis.

\acknowledgments 

We would like to thank Xin Liu, Anna Quider, Simon Dye, Thomas Diehl, and Huan Lin 
for their assistance. We acknowledge Lindsay King for kindly providing the VLT/FORS2 
$R$-band image of the Cosmic Horseshoe. A.E.S. acknowledges support from the 
David and Lucile Packard Foundation and the Alfred P. Sloan Foundation. We wish 
to extend special thanks to those of Hawaiian ancestry on whose sacred mountain 
we are privileged to be guests. Without their generous hospitality, most of the 
observations presented herein would not have been possible.

\bibliographystyle{apj}
\bibliography{apj-jour,lbgrefs,z1sfgrefs}



\end{document}